\documentclass[preprint,aps,prd,showpacs,nofootinbib]{revtex4}

\usepackage{graphicx}
\usepackage{amsfonts,amsfonts,amssymb}

\newcommand*{\chpt}{\raise0.4ex\hbox{$\chi$}PT}
\newcommand*{\schpt}{S\raise0.4ex\hbox{$\chi$}PT}
\def\figref#1{Fig.~\ref{fig:#1}}
\def\Figref#1{Figure~\ref{fig:#1}}

\def\secref#1{Sec.~\ref{sec:#1}}
\def\secrefs#1#2{Secs.~\ref{sec:#1} and \ref{sec:#2}}
\def\Secref#1{Section~\ref{sec:#1}}

\newcommand*{\ie}{\textit{i.e.},\ }
\newcommand*{\eg}{\textit{e.g.},\ }

\def\ala{{\it \`a la}}

\newcommand*{\et}{\textit{et al.}}

\renewcommand*{\prd}[1]{Phys.\ Rev.\ D \textbf{#1}}

\def\tsukubathree{Nucl.\ Phys.\ B (Proc.\ Suppl.) {\bf 129-130} (2004)}

\newcommand*{\MeV}{{\rm Me\!V}}

\newcommand*{\Tr}{\textrm{Tr}}
\newcommand*{\tr}{\textrm{tr}}

\newcommand{\cD}{\mathcal{D}}

\newcommand{\cL}{\mathcal{L}}
\newcommand{\cM}{\mathcal{M}}

\newcommand{\cO}{\mathcal{O}}

\def\eqn#1{\label{eq:#1}}
\def\Equation#1{Equation~(\ref{eq:#1})}

\def\eq#1{Eq.~(\ref{eq:#1})}
\def\eqs#1#2{Eqs.~(\ref{eq:#1}) and (\ref{eq:#2})}

\def\third{{\scriptstyle \raise.15ex\hbox{${1\over3}$}}}
\def\fourth{{\scriptstyle \raise.15ex\hbox{${1\over4}$}}}
\def\bigfourth{\frac{1}{4}}
\def\gtwid{{\,\raise.3ex\hbox{$>$\kern-.75em\lower1ex\hbox{$\sim$}}\,}}
\def\ltwid{{\,\raise.3ex\hbox{$<$\kern-.75em\lower1ex\hbox{$\sim$}}\,}}

\begin{document}

\title{Staggered Chiral Perturbation Theory and the Fourth-Root Trick}
\author{C. Bernard}
\affiliation{Washington University, St. Louis, MO 63130}
\begin{abstract}
Staggered chiral perturbation theory (\schpt) takes into account
the ``fourth-root trick'' for reducing unwanted (taste) degrees of
freedom with staggered quarks by multiplying the contribution
of each sea quark loop by a factor of 1/4. 
In the special case of four staggered fields (four flavors, $n_F=4$), 
I show here that certain assumptions about analyticity and phase structure
imply the validity of this procedure for
representing the rooting trick in the chiral sector.
I start from the observation that, when the four flavors
are degenerate, the fourth root simply reduces $n_F=4$ to $n_F=1$.
One can then treat nondegenerate quark masses by expanding around
the degenerate limit.
With additional assumptions on decoupling, the result can be extended to the more
interesting cases of $n_F=3$, $2$, or $1$.   A apparent paradox associated
with the one-flavor case is resolved.  
Coupled with some expected features of unrooted staggered
quarks in the continuum limit, in particular the restoration of
taste symmetry, \schpt\ then implies that the
fourth-root trick induces no problems (for example, a violation of unitarity
that persists in the continuum limit) in the lowest energy sector of staggered lattice  QCD.
It also says that the theory with staggered valence quarks and rooted
staggered sea quarks behaves like a simple, partially-quenched theory,
not like a ``mixed'' theory in which sea and valence quarks have different
lattice actions.  In most cases, the assumptions made in this paper are
not only sufficient but also necessary for the validity of \schpt, so that
a variety of possible new routes for testing this validity are opened.
\end{abstract}
\maketitle

\section{Introduction}\label{sec:intro}
Staggered quarks \cite{Kogut:1974ag} employ an incomplete reduction
of the lattice doubling symmetry, and therefore have an extra
degree of freedom called ``taste.'' 
In four dimensions, a single staggered field on the lattice produces
four tastes in the continuum limit.  It is possible to interpret taste as physical flavor  ($u$, $d$, $s$, $c$)
by explicitly breaking the continuum taste symmetry with general  mass terms \cite{GENERAL-KS-MASS}.
However, that approach leads to a variety of problems including complex
determinants, violations  of chiral symmetry even in the limit of vanishing light quark masses, 
and the necessity of fine tuning.
The current standard approach --- and the one assumed in this paper ---
is to introduce a separate staggered field for each physical flavor, 
and then attempt to eliminate the unwanted taste degree of freedom by taking
a root of the staggered fermion determinant. 
This procedure was proposed by Marinari, Parisi and Rebbi \cite{Marinari:1981qf} in
a two-dimensional context; a fourth root is required in four dimensions.   Such ``rooted'' staggered quarks have
been used  by the MILC collaboration for recent dynamical simulations \cite{MILC}, 
which give good agreement with experiment for many simple hadronic quantities \cite{PRL}.

There is wide-spread agreement that, whatever their practical problems in reproducing the
desired four-flavor mass spectrum,  ``unrooted'' staggered fermion
quarks are a consistent way to simulate four degenerate tastes of quarks in the continuum limit. 
But the correctness of the fourth root trick to reduce four tastes to one has not been proven,
and there are concerns expressed in the literature about its use
in lattice  QCD simulations \cite{DEGRAND,KENNEDY,DURR}. 
The difficulties arise from the fact
that taste symmetry is broken at order $a^2$,  where $a$ is the lattice spacing.  
This prevents one from implementing the rooting simply by projecting the local four-taste staggered Dirac operator onto a local
operator in a single-taste subspace. Without a local Dirac operator, usual physical properties
of a lattice theory   such as unitarity or universality are called into question.
In the past few years, many authors have 
addressed the issue of the validity of the fourth-root 
procedure \cite{FOURTH-ROOT-TEST-NUMERICAL,ADAMS,SHAMIR,CBMGYS}.  
While a proof is still lacking,
the result of these investigations is to make it rather plausible that staggered
quarks with the fourth-root trick do in fact have the correct continuum limit.
At finite lattice spacing, however, it seems clear that the  fourth root procedure 
introduces a variety of unphysical sicknesses.
This follows not only from the renormalization group 
approach introduced by Shamir \cite{SHAMIR}, but also from the staggered chiral theory, as discussed
below.
The issue is then to prove that these unphysical effects disappear or decouple in
the continuum limit.

Here, I start with a simpler, but related, problem:  What is the chiral theory
that correctly describes rooted staggered quarks?  Lee and Sharpe \cite{LEE-SHARPE}
found  the chiral theory that corresponds to a single unrooted staggered field. In the current
terminology, this is a one-flavor case, with four tastes.
It was generalized to more than one flavor (more than one staggered field) and called
``staggered chiral perturbation theory'' (\schpt) by Aubin and Bernard \cite{AUBIN-BERNARD}.
Certain, rather noncontroversial, assumptions go into these derivations.  In particular,
one needs to know the Symanzik theory \cite{SYMANZIK} that describes 
unrooted staggered quarks as one approaches the
continuum limit.  In deriving the Symanzik theory, one assumes 
that the taste, Lorentz, and translation symmetries become exact in the continuum, and that the
lattice symmetries fit inside the continuum group in a straightforward way.
In addition to theoretical understanding of why this 
should be the case \cite{KS-BASIC,GENERAL-KS-MASS,TASTE-REPRESENTATION,SHAMIR},
there is good numerical evidence for the restoration of these 
symmetries \cite{MILC,FOURTH-ROOT-TEST-NUMERICAL}. 
To find the chiral theory for {\it rooted} staggered quarks, an additional assumption is needed.
In Ref.~\cite{AUBIN-BERNARD}, it was proposed that one could represent the effects of
the fourth root by locating the sea quark loops in \schpt, and then multiplying
each one by a factor of $1/4$.  Here, I take this prescription as defining
what I mean by \schpt\  for rooted staggered quarks.
The question then becomes: Is \schpt\ the correct chiral theory?

In this paper, I show that the validity of
\schpt\ follows from certain nontrivial assumptions on the phase structure
and mass dependence  of the theory.  These assumptions will be
introduced as needed; the most important of them are also collected in
the concluding section. While I will try to argue from
simulations and experience for the plausibility of these assumptions, 
significantly more work is required
to prove and/or numerically test them.  On the other hand,
in most cases it will be clear that the assumptions are not only
sufficient for the validity of \schpt\ but also necessary. Tests of 
the assumptions therefore provide  new means to test 
\schpt\ itself.

Note first of all that \schpt\ for rooted quarks does show unphysical
effects at nonzero lattice spacing.  In the published literature, this
is seen most clearly in Prelovsek's calculation of the flavor
nonsinglet scalar correlator \cite{PRELOVSEK}.  On the lattice, she finds
intermediate-state contributions with mass below that
of the lightest physical intermediate state ($\eta\pi$).  I call this
sickness  a ``unitarity violation'' at finite lattice spacing, since it is
due to contributions from ``extra'' light mesons of various tastes, which only 
cancel completely in the continuum limit.\footnote{One might be 
tempted to describe this sickness
as kind of ``nonlocality'' at finite lattice spacing, because
the correlator decays at long distances at an unphysical rate.  I prefer
to avoid that terminology, because its connection with the standard issue
of the locality of a Dirac operator on the lattice is indirect.}
The flavor-singlet scalar correlator provides
another example of such unitarity violation. It 
has recently been worked out for the three- and one-flavor cases, both unrooted
and rooted \cite{SCALAR}.
Because the one-flavor case is a key test of the ideas discussed
in the current paper, I  present some relevant
details in \secref{one-flavor}.  The scalar correlator at nonzero lattice spacing has
intermediate-state contributions from light pseudo-Goldstone pions, even 
though a one-flavor theory should have only a massive pseudoscalar, the $\eta'$.  
Nevertheless, these
unphysical states decouple from the correlator in the continuum limit.

Thus \schpt\ captures some sicknesses expected  of the rooted theory at nonzero
lattice spacing.  But it is not obvious that \schpt\
captures all such sicknesses. Perhaps there are
other violations of unitarity, or indeed other more subtle features of the rooted theory,
that should be present in the corresponding chiral effective theory but are missed by
\schpt. I argue below that no such effects are missed.
 The starting point is a special
case in which there is virtually no doubt about the correct chiral theory: a
rooted theory with four degenerate quark flavors. In this case, the fourth-roots
of the four determinants are identical, so their product just gives the determinant
of a single, unrooted staggered field. (Note that the staggered determinant
is positive, and the algorithmic treatment of the rooting trick in the simulations 
gives the positive fourth root \cite{ALGORITHM}.) With the noncontroversial assumptions
mentioned above, the corresponding chiral theory is just the \schpt\ of Lee and Sharpe
\cite{LEE-SHARPE}.  

One can then expand around the degenerate case
to treat the case of nondegenerate masses.  For technical reasons,
this requires the use of a partially quenched chiral theory with valence masses
degenerate with those of the sea quarks. Golterman, Sharpe and Singleton (GSS) \cite{GSS}
show that the phase structure of a quenched chiral theory can be subtle,
and analogous questions can be raised about the partially quenched
theory.  
The use of partial quenching in  this paper seems to be safe from
any GSS subtleties. However, since the theory has not yet been investigated in detail using
the GSS methods, I highlight a few places where complications could
conceivably enter.  Further investigation along the lines of Ref.~\cite{GSS} is planned.

The completion of the argument for four nondegenerate flavors 
requires nontrivial assumptions about the analytic structure of the mass dependence.  
In particular, I need to assume that
there is no
essential singularity at zero degeneracy in the difference between \schpt\ and the putative
correct chiral theory. Phase transitions in the chiral theory at nonzero quark mass
differences  would also be dangerous, although the 
existing simulations \cite{MILC} can be put forward as evidence against
such phase transitions, at least in the region of parameter space investigated to date.

To move to the phenomenologically more interesting case of three light flavors,
the mass of one quark can be taken large.  In \schpt, it is quite clear that
the heavy quark will decouple, leaving three-flavor \schpt. However, in the lattice  QCD of
rooted staggered quarks, the nature of the decoupling, while in my opinion plausible, 
requires an additional assumption.  With this assumption, it follows that \schpt\ is the
correct chiral description of the rooted three flavor theory. The process can
then be repeated, leading to statements about the two- and the one-flavor theories.

If \schpt\ is accepted as the correct chiral description, it provides
strong evidence that rooted staggered quarks have the desired continuum limit,
in other words that they are in the correct universality class.
The point is that \schpt\ automatically becomes continuum chiral perturbation theory (\chpt)
in the continuum limit, modulo the usual assumptions on
the restoration of taste symmetry in the continuum limit of unrooted staggered
quarks.  Therefore, this line of reasoning says that the low energy
(pseudoscalar meson) sector of  lattice QCD with rooted staggered quarks is, in the
continuum limit, indistinguishable from
that of ordinary QCD.   This would significantly reduce the ``phase space'' for any
possible sicknesses of rooted staggered quarks in the continuum limit.

Another consequence of the arguments in this paper is more technical:
If \schpt\ is valid, the lattice theory
with rooted staggered sea quarks and ordinary staggered valence quarks (the
theory in the MILC simulations \cite{MILC}) behaves like
a ``partially quenched'' theory.\footnote{In the continuum limit, this was anticipated in
Ref.~\protect{\cite{PQ}}.} Effectively, this means that there are symmetries that connect
valence and sea quarks.  As usual for a partially quenched theory, such symmetries
may be broken in a controlled way by mass differences between valence
and sea quarks. However, the symmetries are not broken by lattice
corrections. The theory therefore does not behave like a ``mixed'' theory, in which
valence and sea quarks have different lattice actions.  In the mixed case, there
are no symmetries at finite lattice spacing that connect valence and sea quarks.
The chiral descriptions of mixed theories \cite{MIXED} thus have terms --- vanishing in
the continuum limit --- that violate such symmetries. These terms can, for example,
lead to mass splittings between mesons composed of  two valence quarks and those
composed of one valence and one sea quark.  
I show here that the chiral theory
for staggered valence and rooted staggered sea quarks does {\it not}\/ have such terms;
corresponding valence-valence, valence-sea, and sea-sea mesons are degenerate. 

The remainder of this paper is organized as follows:  In \secref{replica},
I discuss the replica trick in \schpt; this is a 
systematic way to find sea quark loops in the chiral theory and
multiply each by a factor $1/4$. \Secref{notation} then introduces the notation
needed to describe the various theories considered here, at both the
QCD and the chiral levels, and makes some introductory comments about these theories.
The details of my assumptions and arguments for \schpt\ in the four-flavor case are presented
in \secref{details}; while the extension to three or fewer flavors is treated
in \secref{extension}.  
\Secref{one-flavor} shows in some detail
how the one-flavor case works.  I resolve there the apparent paradox of light pseudo-Goldstone
mesons appearing the one-flavor chiral theory.
Consequences of my arguments for the rooted theory at the QCD level are
described in \secref{consequences}. 
Finally, I review the  assumptions and conclusions and make some additional remarks 
and speculations in \secref{conclusions}.

\section{Replica Trick}
\label{sec:replica}

In \schpt\ for rooted staggered quarks, one needs to identify the presence of sea quark loops
in various meson diagrams, and multiply each such loop by a factor of $1/4$.
The sea quark loops were located  in Ref.~\cite{AUBIN-BERNARD} by  using the quark
flow approach \cite{QUARK-FLOW}. While quark flow gives a rather intuitive physical picture,
it suffers from the disadvantage in the current case that it is formulated as a series of
rules for tracing flavor indices, not as an algebraic statement.
The replica trick provides an alternative approach that is systematic and algebraic. 
It was applied to partially quenched theories by
Damgaard and Splittorff \cite{REPLICA} and was first used for \schpt\ in Ref.~\cite{AUBIN-BERNARD-REPLICA}.

The replica procedure for rooted \schpt\ is very simple:  
One starts by replicating the sea-quark flavors,
replacing each dynamical staggered field by $n_R$ identical copies, where $n_R$ is a
(positive) integer. One then calculates straightforwardly order by order in 
the corresponding (unrooted) \schpt, keeping the $n_R$ dependence explicit.  
Finally, one sets $n_R=1/4$.  

Note that, at any finite order in \schpt, the $n_R$ dependence is polynomial: It just
comes from the sum over the sea quark indices in meson loops.  
Therefore, the process of taking $n_R\to \fourth$ is straightforward and unambiguous
order by order.  
As always in chiral perturbation theory, we treat the low energy constants
(LECs) as free parameters for each $n_R$. We should not use any relations
that hold only for special values of $n_R$ --- analogous to those
discussed by Sharpe and Van de Water \protect{\cite{Sharpe:2003vy}} ---
to reduce the number of chiral operators. If it turns out that we are left with some
redundant operators when $n_R\to\fourth$, we can always redefine the LECs to absorb
the redundancy at the end.
Within chiral perturbation theory, we do not worry about 
(nor do we have any control over) the dependence of low energy constants themselves on
$n_R$.  Such dependence, coming from an underlying QCD-like theory, would in fact
be nonperturbative in the strong coupling $\alpha_S$
and probably not polynomial in $n_R$.

At the QCD level, it is difficult to give the replica trick 
any meaning beyond weak-coupling perturbation theory, in which
the $n_R$ dependence is again polynomial. Within weak-coupling perturbation theory,
the replica trick is in fact somewhat useful, because it provides a convenient way of keeping track of
sea-quark loops. This can aid in clarifying
the argument in  Ref.~\protect{\cite{PQ}} of the validity of the fourth-root procedure in perturbation theory,
and will also  be  helpful in \secref{mixed}.
Nonperturbatively, however,
even if we were to assume that the $n_R\to\fourth$ limit should be
taken by analytic continuation, the replica trick would be ambiguous since there
is no unique continuation from the integers.

A related comment is that 
the use of the replica trick for a chiral theory
is valid, {\it a priori}\/, only for order by order 
calculations in  chiral perturbation theory. We have no
guarantee of its correctness in general nonperturbative chiral calculations, such as
the determination of the correct vacuum state. 
However, in the 
degenerate four-flavor theory, we know the chiral theory (and hence
the appropriate phase) independent of the replica trick. 
As I move away from the degenerate limit, I will in any case need
to assume that dependence on quark mass is smooth and no phase change occurs (see \secref{details}).
Thus, there is no further restriction coming from the perturbative
nature of the replica trick.

\section{Theories Considered; Notation}
\label{sec:notation}
We need some notation to refer to the various versions of QCD and their
corresponding chiral theories.
Define a version of lattice QCD by $(n_F,n_T,n_R)_{LQCD}$, where $n_F$ is the number
of flavors (the number of staggered fields),  $n_T$ is the number
of tastes per field, and $n_R$ is the number
of replicas.  
The corresponding chiral theories are denoted by  $(n_F,n_T,n_R)_{\chi}$.
If $n_R$ is trivially equal to 1
(because the replica trick is not relevant), it is omitted. 

When $(n_F,n_T,n_R)_{\chi}$ or  $(n_F,n_T,n_R)_{LQCD}$ are used in equations, 
I am referring specifically to the generating functionals for these theories, with sources to be discussed below.

I focus primarily on three versions of QCD, and four versions of chiral theories:
\begin{itemize}

\item{}  $(1,4)_{\chi}$ and $(1,4)_{LQCD}$: These are the chiral and QCD theories of a single staggered field
(one flavor) with four tastes.  By (noncontroversial) assumption, the chiral theory $(1,4)_{\chi}$ is just the
\schpt\ of Lee and Sharpe \cite{LEE-SHARPE}.  No rooting is done at the QCD level, and no
replica trick is necessary at the chiral level.

\item{}  $(n_F,4,n_R)_{\chi}$ and $(n_F,4,n_R)_{LQCD}$: These are the theories for $n_F$ staggered fields
($n_F$ flavors), each replicated $n_R$ times. When $n_R$ is indicated explicitly,
as in this case, it is taken to be an integer only; no rooting is done. The 
chiral theories $(n_F,4,n_R)_{\chi}$ are  --- again by noncontroversial assumption --- just those
of Aubin and Bernard \cite{AUBIN-BERNARD} for integer ($n_F\cdot n_R$) number of flavors.
They are obtained from the $n_F$ flavor chiral theories by
replicating the sea-quark degrees of freedom in the chiral fields.

\item{} $(n_F,``1")_{\chi}$ and $(n_F,``1")_{LQCD}$: These are the chiral and QCD theories of $n_F$ staggered fields
($n_F$ flavors) with the  $\root 4 \of {\rm Det}$ taken at the QCD level to reduce 4 tastes to 1 for each flavor.  
Since I do not want to assume here that the rooting procedure is correct, I write the $1$ for tastes
in quotation marks.   Then $(n_F,``1")_{\chi}$ is by definition the chiral theory generated by  $(n_F,``1")_{LQCD}$.
The main point of this paper is to construct 
$(n_F,``1")_{\chi}$ unambiguously.

\item{} $(n_F,4,\fourth)_\chi$: This is the chiral theory $(n_F,4,n_R)_{\chi}$, now implementing 
the replica trick  
by taking $n_R\to \fourth$, with the goal of describing
rooted staggered quarks.  
In the literature
(\eg Ref.~\cite{AUBIN-BERNARD,AUBIN-BERNARD-REPLICA,HEAVYLIGHT,SCHPT-OTHER,SCHPT-BARYONS}), 
it is {\it assumed}\/ that 
this procedure produces the right chiral theory; in other words, it is assumed
that $(n_F,``1")_{\chi}$ = $(n_F,4,\fourth)_\chi$. 
Here, I define \schpt\ as $(n_F,4,\fourth)_\chi$, 
and then ask  
the question of whether \schpt\ is indeed the correct chiral theory.
Note that I avoid reference to corresponding QCD theories ``$(n_F,4,\fourth)_{LQCD}$'' because,
as discussed in \secref{replica}, I do not know how to
give unambiguous meaning beyond perturbation theory to the replica trick for those QCD-level theories.

\end{itemize}

For my arguments, the chiral theories $(n_F,4,n_R)_{\chi}$ are key objects.  
On the other hand, the corresponding QCD theories
 $(n_F,4,n_R)_{LQCD}$, in particular  $(4,4,n_R)_{LQCD}$,
are introduced for convenience,  because they allow one to keep
track more easily of the factors of $n_R$ that relate valence- to sea-quark matrix elements (see
\secref{details}).  These QCD-level theories
can be eliminated at the expense of a somewhat less intuitive argument  at the chiral level, related to quark flow.
An outline of such an alternative argument
 is given in \secref{results-nf4}; it does however seem to require
a weak additional assumption.  Because the   $(4,4,n_R)_{LQCD}$ theories are just used formally,
it is probably unnecessary that the standard, broken realization 
of chiral symmetry assumed in $(4,4,n_R)_{\chi}$ actually occurs in $(4,4,n_R)_{LQCD}$. 
The unpleasant fact that
asymptotic freedom (and presumably spontaneous chiral symmetry breaking) is lost for  $n_R>1$ 
in $(4,4,n_R)_{LQCD}$ seems to be irrelevant. An easy way to see this is to realize that
the precise correspondence between $(4,4,n_R)_{LQCD}$ and $(4,4,n_R)_{\chi}$ can be 
maintained by an artifice,\footnote{I thank
Urs Heller for this comment.} as follows:
Note first that the order of the polynomial
dependence on $n_R$ is bounded at a given order in chiral perturbation theory. This means there is maximum
value of $n_R$, $n_R^{\rm max}$, that need be considered in order to determine the polynomial completely.
One can then simply imagine increasing the number of colors sufficiently
to ensure that the QCD theory has the standard, spontaneously broken,
realization of chiral symmetry
for any $n_R \le n_R^{\rm max}$.  Recall that
the mesonic chiral theory generated by a given $(4,4,n_R)_{LQCD}$ is independent of the number
of colors as long as the phase is unchanged.  The numerical values of the LECs do depend
on the number of colors, but we are uninterested in those values here.

In the next section, I argue that the replica trick produces the correct chiral theory in the four-flavor case.  
In other words, I claim that
\begin{equation}\eqn{toshow}
(4,``1")_\chi \doteq (4,4,\fourth)_\chi \ .
\end{equation}
This should be taken as a statement about the generating functionals of the two chiral theories.
I use ``$\doteq$,'' rather than ``$=$,'' to compare  two chiral theories, because what I 
mean is that they are the same functions of the LECs: True equality would only result if we
adjusted the  LECs to be the same.

One also needs to be careful about what sources (equivalently, external fields)
one is allowing in the
Green's functions on both sides of equations such as \eq{toshow}.  For example, there are more sea-quark fields
available in the $(4,4,n_R)_\chi$ theory, from which $(4,4,\fourth)_\chi$ is obtained, than there
are in the $(4,``1")_\chi$ theory.
Unless explicitly stated otherwise, such generating functionals should be taken to
describe partially quenched theories, with sources coupled to valence fields
only.  Ghost (bosonic) fields,  degenerate with the valence fields
and required to cancel the valence determinant,
are also implicit. 
When I need to make the sources explicit, I will include any
valence sources $\sigma$ among the arguments, for example $(n_F,n_T,n_R;\sigma)_{LQCD}$.
Identical staggered valence fields with identical
valence sources are always assumed on both sides of equations relating generating functionals. 

\section{Details of the Argument for Four Flavors}
\label{sec:details}

The key ingredient is the observation that, when there are
four degenerate flavors (four staggered fields with equal masses), the rooting procedure 
clearly reduces the four flavor
theory to a one flavor theory. In other words, instead of acting on tastes
and (presumably) reducing the four tastes per flavor to one taste per flavor,
we can think of the rooting in this case as acting on flavor and reducing
four fields to one, without affecting the tastes. 
Let the quark mass matrix be $\cM$. The condition
of degeneracy is $\cM = \bar m I$, where $\bar m$ is a number and $I$ is the
identity matrix in flavor space. It then follows that:
\begin{eqnarray}\eqn{deg-root-QCD}
(4,``1")_{LQCD}\Big\vert_{\cM=\bar m I} & = & (1,4)_{LQCD}\Big\vert_{\bar m} \\
\eqn{deg-root-chi}
(4,``1")_{\chi}\Big\vert_{\cM=\bar m I} & \doteq & (1,4)_{\chi}\Big\vert_{\bar m} \ \doteq\  (4,4,\fourth)_\chi\Big\vert_{\cM=\bar m I} \ .
\end{eqnarray}
The last equivalence in \eq{deg-root-chi} is manifest order by order in \schpt: Since the
result for any physical quantity is polynomial in the number of degenerate flavors, taking
$4n_R$ degenerate flavors and then putting $n_R=1/4$ gives the same chiral expansion
as a one-flavor theory.

One can make a stronger statement than \eq{deg-root-chi} by adding
sources and computing
specific Green's functions in the degenerate case.  In order to keep the
arguments simple, I generally use only taste-singlet scalar sources, which are all
that are necessary to allow us to move beyond the degenerate mass limit. For
writing explicit terms in the chiral theory, however, it will be convenient below to include 
pseudoscalar sources temporarily, since it is linear combinations of
scalar and pseudoscalar source that transform simply under chiral
transformations. One can also easily generalize to
sources of arbitrary taste if desired.

I start by adding
introducing the scalar sources into the sea-quark sector of the QCD-level theory
$(4,``1")_{LQCD}$.
Let
$\Psi_i(x)$ be the sea quark field of flavor $i$ at space-time point $x$. For convenience,
I work in the taste 
representation \cite{TASTE-REPRESENTATION}, with taste (and spin) indices on $\Psi$
implicit, but there is no reason why one cannot work directly
with the one-component staggered fields instead. 
The source $s(x)$ is taken to be a Hermitian matrix in flavor space.
The mass and source terms are then:
\begin{equation}\eqn{source-41}
\bar m\, \bar \Psi_i(x) \Psi_i(x) + \bar \Psi_i(x) \, s^{ij}(x)\, 
\Psi_j(x)\;, \hspace{1.5cm}(4,``1")\ {\rm case},
\end{equation}
where sums over flavor indices $i,j$ are implied.

One needs to state precisely here what is meant by
a  rooted staggered theory with sources.  In this paper, I always mean:
(1) introduce the sources into the corresponding unrooted theory;
(2) integrate the sea quark
fields to get a determinant that is a function of the sources; 
(3) replace the determinant by its fourth root.  
Derivatives with respect to the sources, if desired, are taken only after step (3).

Now introduce the same  sources into the replica QCD theories
$(4,4,n_R)_{LQCD}$, 
with the specification that a given source couples equally to 
all replicas. 
We have:
\begin{equation}\eqn{source-44nR}
\bar m \bar \Psi^r_i(x) \Psi^r_i(x) + \bar \Psi^r_i(x) \,s^{ij}(x)
\, \Psi^r_j(x)\;, \hspace{1.5cm}(4,4,n_R)\ {\rm case} .
\end{equation}
Sums over the replica index $r = 1,2,\dots,n_R$, as well as the flavor indices $i$ and $j$, are implied.

When the sources are nonzero (which includes the case of
nondegenerate quark masses as a special case), we do not yet know that
$(4,4,\fourth)_\chi$ is the right chiral theory. One could imagine that there
are extra terms in $(4,``1")_\chi$ that vanish in the limit
$s=0$. So I define the difference to be an unknown functional $V[s]$:
\begin{equation}\eqn{correction}
(4,``1";\,s)_\chi \doteq (4,4,\fourth;\,s)_\chi + V[s] \ ,
\end{equation}
where $V[0]\!=\!0$.  As far as we know at this point, $V[s]$ could be quite sick.
For example, it could generate Euclidean correlation functions with unphysical decay rates
(unphysical intermediate states), even in the continuum limit.

There are further restrictions on $V[s]$ coming from
the fact that the two chiral theories must be equivalent when there is
exact flavor symmetry.
We must have $V[s]\!=\!0$ whenever $s(x)$ is proportional to the identity in flavor
space or can be brought there by an
$SU(4)_L\times SU(4)_R$  chiral flavor rotation. 
Therefore it takes some care even to write down a possible term in $V[s]$.

I temporarily add a Hermitian pseudoscalar source $p(x)$ to the theories. For example,
corresponding to \eq{source-41} is
\begin{equation}\eqn{psource-41}
 i\bar \Psi_i(x) \, \gamma_5\, p^{ij}(x)\, 
\Psi_j(x)\;, \hspace{1.5cm}(4,``1")\ {\rm case}.
\end{equation}
The spurion combinations 
$h \equiv \bar mI+ s +ip$ and $h^\dagger \equiv \bar mI+ s-ip$ transform
simply under chiral rotations $L\in SU(4)_L$ and $R\in SU(4)_R$:
\begin{equation}\eqn{h-transform}
h \to L\,  h\, R^\dagger\ , \hspace{1.5truecm}  h^\dagger\to R\, h^\dagger\, L^\dagger \ .
\end{equation}
If
\begin{equation}\eqn{deg-condition}
 h(x) = c(x) U \ , \hspace{1.5truecm} h^\dagger(x) = c^*(x) U^\dagger \ ,
\end{equation}
where $U\in SU(4)$ is a constant matrix and $c(x)$ is a c-number function, then 
$h(x)$ and 
$h^\dagger(x)$ can be made everywhere proportional to the identity by the chiral rotation
$R=U$, $L=I$ and there is exact
flavor symmetry, unbroken by masses or sources.

We can now look for possible terms in $V$, at first expressed 
as functionals of $h$ and $h^\dagger$.
An example that satisfies the above requirements 
is
\begin{equation}
\tilde V_1 = \int d^4x\, d^4y
\left(\frac{1}{\square + M^2}
\right)_{\hspace{-.13cm}x,y} \hspace{-.08cm}
\bigg( \Tr\left[ h(x)\, h^\dagger(x)\, h(y)\, h^\dagger(y)\right] 
- \fourth\Tr\left[ h(x)\, h^\dagger(x)\right]
	\Tr\left[ h(y)\, h^\dagger(y)\right] \bigg)
\eqn{v1big}
\end{equation}
where $\Tr$ is a flavor trace, 
and $1/M$ a distance scale that might not go
to zero in the continuum.  For example, one could have
$M=k\Lambda_{QCD}$, where $k$ is some constant. In the worst case, $M$ might not 
even correspond to the mass of any physical particle in QCD.

Removing the pseudoscalar source $p(x)$ and keeping only the lowest nonvanishing term in $s$, one gets the following
example of a possible contribution to $V[s]$:
\begin{equation}
V_1 = 4\bar m^2\int d^4x\; d^4y \;
\left(\frac{1}{\square + M^2}
\right)_{x,y} 
\bigg( \Tr\left[ s(x)s(y)\right] 
- \fourth\Tr\left[ s(x)\right]\,
	\Tr\left[ s(y)\right] \bigg)
\eqn{v1}
\end{equation}
The goal is of course to prove that $ V[s]$ actually vanishes.

\subsection{Expansion around the Degenerate Theory}
\label{sec:expansion}
If we take derivatives of the generating functionals 
with respect to $s$ and evaluate them
at $s=0$, we will have Green's functions for degenerate quark masses.
At  the level of the chiral theories, I claim that \eqs{deg-root-QCD}{deg-root-chi} 
(modulo some technical assumptions) actually imply the stronger statement:
\begin{equation}\eqn{derivs}
\prod_n\frac{\partial}{\partial s^{i_nj_n}(x_n)} 
(4,``1";\,s)_\chi\Big\vert_{s=0}\doteq\;\;
\prod_n\frac{\partial}{\partial s^{i_nj_n}(x_n)} 
(4,4,\fourth;\,s)_\chi\Big\vert_{s=0}  
\end{equation}
for any given combination of derivatives with respect to $s$.

A difficulty in proving
\eq{derivs} is that, as soon as  the sources are taken
to be nonzero in order to compute the derivatives, we no longer know
that $(4,``1")_\chi$ and $(4,4,\fourth)_\chi$ are the same.
Further, I must avoid the use of $(4,4,\fourth)_{LQCD}$, which is not well defined.
Finally, I cannot use $(1,4)_{LQCD}$ 
easily as an intermediate step, because sea quark sources with nontrivial
flavor ($s^{ij}$) cannot be inserted into a one-flavor theory.

The need for nonzero sea-quark sources in \eq{derivs} can be circumvented
by using valence sectors, in other words, by considering the partially quenched version
of \eqs{deg-root-QCD}{deg-root-chi}.  I thus introduce into all theories of interest
an arbitrary number, $n_V$, of staggered valence fields
$q_\alpha$, where $\alpha= 1,2,\dots n_V$ is the valence flavor index.
These have degenerate mass $\bar m$ and are coupled to valence sources $\sigma_{\alpha\beta}$, 
giving mass and source terms as follows:
\begin{equation}\eqn{val-source}
\bar m \bar q_\alpha(x) q_\alpha(x) + \bar q_\alpha(x) \, \sigma^{\alpha\beta}(x)\, 
 q_\beta(x) \ ,
\end{equation}
with sums over $\alpha$ and $\beta$ implied.  The valence-quark 
source $\sigma^{\alpha\beta}$ is
exactly analogous to the sea-quark source $s^{ij}$; they only differ in the type of quarks to which they
couple.

I also introduce $n_V$ corresponding ghost (bosonic) quarks, again with
degenerate mass $\bar m$.  These ghosts do not couple to the $\sigma^{\alpha\beta}$ source,
so that derivatives with respect to $\sigma^{\alpha\beta}$ produce Green's functions made 
purely of (fermionic) valence quarks.   When $\sigma^{\alpha\beta}=0$, the valence
and ghost determinants cancel.

The partially quenched version of \eq{deg-root-QCD} remains valid,
since the valence/ghost sectors are identical on both sides, and the sea-quark determinants are equal
as long as the sea-quark source $s$ vanishes (giving degenerate masses):
\begin{equation}\eqn{deg-root-PQQCD}
(4,``1";\,s\!=\!0,\sigma)_{LQCD}  \hspace{0.2cm}=\hspace{0.2cm} (1,4;\,s\!=\!0,\sigma)_{LQCD} \ ,
\end{equation}
where sea and valence sources are indicated explicitly.
  
The equality of generating functionals must also be true for the corresponding chiral theories:
\begin{equation}\eqn{deg-root-PQchi}
(4,``1";\,s\!=\!0,\sigma)_{\chi} \hspace{0.2cm} \doteq\hspace{0.2cm} (1,4;\,s\!=\!0,\sigma)_{\chi} \ ,
\end{equation}
This follows by definition of what it means to be the corresponding chiral theory.  
I am assuming that such partially quenched chiral theories exist. But note
that the starting LQCD theories both have local actions, so this appears
to be a rather safe assumption.  I am not
claiming, however, that I know explicitly how to calculate ghost or valence 
Green's functions in either of these
chiral theories.  My expectation is that the ``naive'' meson Feynman rules, which follow from
the methods of Ref.~\cite{PQ}, are probably correct.  However,
to prove that would require an analysis along the lines of Ref.~\cite{GSS} to determine
the proper saddle point for the mesons constructed from valence or ghost quarks, 
around which the chiral perturbation
theory can be developed.  Such an analysis is in progress.

In discussing \eq{deg-root-chi}, I claimed that the equivalence 
of the $(1,4)_\chi$ and $(4,4,\fourth)_\chi$ theories is
``manifest'' order by order in \schpt.   In the presence of
valence/ghost fields and sources, the corresponding statement is almost certainly still true.
Even if the saddle point for ghost mesons (or valence) mesons is nontrivial, it 
is very difficult to see how it could be affected, order by order,  by the difference 
between having one sea-quark flavor
or having $4n_R$ degenerate sea flavors and then putting $n_R=1/4$.  
Combined with \eq{deg-root-PQchi}, this gives
\begin{equation}\eqn{deg-replica-PQchi}
(4,``1";\,s\!=\!0,\sigma)_{\chi} \hspace{0.2cm} \doteq\hspace{0.2cm} (4,4,\fourth;\,s\!=\!0,\sigma)_{\chi} \ .
\end{equation}

In the limit $s=0=\sigma$, all quarks, both valence and sea,
are degenerate. This means one can relate Green's functions constructed from sea-quark fields
to those constructed from valence fields, or equivalently, relate derivatives with respect
to  $s$ to those with respect to $\sigma$.  This is not
completely straightforward, however.  In the (4,``1") theory,
derivatives with respect to $s$
bring down factors of $1/4$ from $\root 4 \of {{\rm Det}(D+\bar m + s )}= \exp \fourth \tr 
\ln (D+\bar m + s ) $. When more than one contraction (more than one
term resulting from the derivatives) is possible, different contractions will be associated
with different numbers of factors of $1/4$.  The power of $1/4$ is just the
number of quark loops implied by the contractions.    

On the other hand, with arbitrary
$n_V$, we can always adjust the flavors of the valence sources being differentiated
so that only one contraction is possible.  This means we can always write an arbitrary derivative
of the generating functional with respect to $s$ as a linear combination
of derivatives with respect to $\sigma$, each term being multiplied by 
$(\fourth)^L$, where $L$ is the number of valence loops in the term. 
The following two examples should clarify what I mean; take flavors $i\not=j$ and $\alpha\not=\beta$
and do not sum over repeated indices:
\begin{eqnarray}
\eqn{example-ij}
\frac{\partial}{\partial s^{ij}(x)} 
\frac{\partial}{\partial s^{ji}(y)} 
\; (4,``1";\,s,\sigma\!=\!0)_{LQCD}\Big\vert_{s=0}&=& 
-\bigfourth\;\langle \tr\Big( G_{j}(x,y)  G_{i}(y,x) \Big) \rangle\nonumber  \\
&&\hspace{-3.0truecm}
=\bigfourth\;\frac{\partial}{\partial \sigma^{\alpha\beta}(x)} 
\frac{\partial}{\partial \sigma^{\beta\alpha}(y)} 
\; (4,``1";\,s\!=\!0,\sigma)_{LQCD}\Big\vert_{\sigma=0} \\ 
\frac{\partial}{\partial s^{ii}(x)}
\frac{\partial}{\partial s^{ii}(x)}
\; (4,``1";\,s,\sigma\!=\!0)_{LQCD}\Big\vert_{s=0}=\nonumber && \\
&&\hspace{-6.0truecm}
=-\bigfourth\;\langle \tr\Big( G_{i}(x,y)  G_{i}(y,x) \Big)\rangle + 
\left(\bigfourth\right)^2\langle \tr\Big( G_{i}(x,x)\Big) \tr\Big(  G_{i}(y,y) \Big) \rangle\nonumber \\
&&\hspace{-6.0truecm}
=\left[\bigfourth\;\frac{\partial}{\partial \sigma^{\alpha\beta}(x)}
\frac{\partial}{\partial \sigma^{\beta\alpha}(y)} +
\left(\bigfourth\right)^2\frac{\partial}{\partial \sigma^{\alpha\alpha}(x)}\frac{\partial}{\partial \sigma^{\beta\beta}(y)}
\right]
(4,``1";\,s\!=\!0,\sigma)_{LQCD}\Big\vert_{\sigma=0} \nonumber \\
&&
\eqn{example-ii}
\end{eqnarray}
where $G_{i}(y,x)$ is the propagator of a quark of flavor $i$ from $x$ to $y$, expectation
values are taken in the $(4,``1")_{LQCD}$ theory with $\cM=\bar m I$ and vanishing sources, 
and the traces are over taste and spin indices.  Note that the two sides of \eq{example-ij}
or \eq{example-ii} are just two different ways of expressing the expectation value of the
same combination of quark propagators, so no subtleties of partial quenching 
\ala\/\ Ref.~\cite{GSS} can
interfere with the equality.

With enough derivatives with respect to $s$, there will always be enough
repeats in sea quark flavor indices that more than one contraction contributes. On the other hand,
since we have an arbitrary number of valence quarks at our disposal, we can always arrange
the valence flavors in the derivatives with respect to $\sigma$ so that only
one contraction occurs.

In the $(4,4,n_R)_{LQCD}$ theory, equations very similar to \eqs{example-ij}{example-ii} hold, with
the simple replacement $\fourth\to n_R$. The factors of $n_R$
are produced by the sum over replicas for each quark loop. 

For an arbitrary $k^{\rm th}$ derivative of  $(4,``1")_{LQCD}$ or $(4,4,n_R)_{LQCD}$ with respect to
$s$, we therefore can write:
\begin{eqnarray}\eqn{QCD-derivs-41}
\prod_{n=1}^k\frac{\partial}{\partial s^{i_nj_n}(x_n)} 
\, (4,``1";\,s,\sigma\!=\!0)_{LQCD}\Big\vert_{s=0}= &&\nonumber \\
&&\hspace{-3.5cm} =\sum_C \left(\bigfourth\right)^{L_C}
\prod_{n=1}^k\frac{\partial}{\partial \sigma^{\alpha^C_{n}\beta^C_{n}}(x_n)} 
\, (4,``1";s\!=\!0,\sigma)_{LQCD}\Big\vert_{\sigma=0} \\ 
\prod_{n=1}^k\frac{\partial}{\partial s^{i_nj_n}(x_n)} 
\, (4,4,n_R;\,s,\sigma\!=\!0)_{LQCD}\Big\vert_{s=0}= &&\nonumber \\
&&\hspace{-3.5cm} =\sum_C \left(n_R\right)^{L_C}
\prod_{n=1}^k\frac{\partial}{\partial \sigma^{\alpha^C_{n}\beta^C_{n}}(x_n)} 
\, (4,4,n_R;\,s\!=\!0,\sigma)_{LQCD}\Big\vert_{\sigma=0} 
\eqn{QCD-derivs-44nR}
\end{eqnarray}
where $C$ labels a particular contraction with $L_C$ valence quark loops, and the valence flavor
indices $\alpha^C_{n}$ and $\beta^C_{n}$
are adjusted
to make only that contraction possible.  The key point in \eqs{QCD-derivs-41}{QCD-derivs-44nR}
is that the same arrangements of valence flavor indices and powers $L_C$ work in both cases.

We now pass to the chiral theory in both cases, giving:
\begin{eqnarray}\eqn{chi-derivs-41}
\prod_{n=1}^k\frac{\partial}{\partial s^{i_nj_n}(x_n)} 
\, (4,``1";\,s,\sigma\!=\!0)_{\chi}\Big\vert_{s=0}= &&\nonumber \\
&&\hspace{-2.8cm} =\sum_C \left(\bigfourth\right)^{L_C}
\prod_{n=1}^k\frac{\partial}{\partial \sigma^{\alpha^C_{n}\beta^C_{n}}(x_n)} 
\, (4,``1";s\!=\!0,\sigma)_{\chi}\Big\vert_{\sigma=0} \\ 
\prod_{n=1}^k\frac{\partial}{\partial s^{i_nj_n}(x_n)} 
\, (4,4,n_R;\,s,\sigma\!=\!0)_{\chi}\Big\vert_{s=0}= &&\nonumber \\
&&\hspace{-2.8cm} =\sum_C \left(n_R\right)^{L_C}
\prod_{n=1}^k\frac{\partial}{\partial \sigma^{\alpha^C_{n}\beta^C_{n}}(x_n)} 
\, (4,4,n_R;\,s\!=\!0,\sigma)_{\chi}\Big\vert_{\sigma=0} 
\eqn{chi-derivs-44nR}
\end{eqnarray}
At any finite order in chiral perturbation theory, both sides of 
\eq{chi-derivs-44nR} are polynomial in $n_R$.  Therefore 
the limit $n_R\to\fourth$ is well defined:
\begin{eqnarray}
\prod_{n=1}^k\frac{\partial}{\partial s^{i_nj_n}(x_n)} 
\, (4,4,\fourth;\,s,\sigma\!=\!0)_{\chi}\Big\vert_{s=0}= &&\nonumber \\
&&\hspace{-2.8cm} =\sum_C \left(\bigfourth\right)^{L_C}
\prod_{n=1}^k\frac{\partial}{\partial \sigma^{\alpha^C_{n}\beta^C_{n}}(x_n)} 
\, (4,4,\fourth;\,s\!=\!0,\sigma)_{\chi}\Big\vert_{\sigma=0} 
\eqn{chi-derivs-4414}
\end{eqnarray}

The right-hand sides
of \eqs{chi-derivs-41}{chi-derivs-4414} are now equal by \eq{deg-replica-PQchi}.
On the left-hand sides, the valence and ghost contributions cancel completely
since $\sigma=0$, so we may eliminate those fields.
This proves \eq{derivs}.

\subsection{Assumptions and Results in the Four Flavor Theory}
\label{sec:results-nf4}
\Equation{derivs}, together with the definition of $V[s]$, \eq{correction}, imply that
$V[s]$ and all of its derivatives vanish at $s=0$:
\begin{equation}\eqn{Vs-derivs}
\prod_{n=1}^k\left(\frac{\partial}{\partial s^{i_nj_n}(x_n)} 
\, V[s]\right)\Bigg\vert_{s=0} = 0\ .
\end{equation}
Thus terms like $V_1$ in \eq{v1} are ruled out. 
Indeed, if $V[s]$ is assumed to be an analytic function,\footnote{At this
point it is sufficient for my purposes to restrict $s$ to a constant matrix, just giving the mass differences.
Therefore $V$ can be thought of as a function, not a functional, and there is no subtlety
with concepts such as analyticity.} with any number of isolated singularities,
it follows that $V[s]\!=\!0$ everywhere. In other words, 
\eq{toshow} is true under this assumption.   
Normally one expects that when
a function is expanded in a Taylor series around some point, the expansion will have
a finite radius of convergence, given by the location of the closest singularity.  But here,
every term in the expansion is zero, so we can continue past any purported isolated singularity,
and thereby show that the singularity is actually absent.

Note that $(4,4,\fourth)_\chi$, as a limit of the replica theories when $n_R\to\fourth$,
is only defined order by order in chiral perturbation theory. By definition,
therefore, the vacuum state of $(4,4,\fourth)_\chi$  has the standard broken realization 
of chiral symmetry that appears in $(4,4,n_R)_\chi$.
We know this is the correct nonperturbative vacuum in the degenerate limit, because there one
can use the chiral theory $(1,4)_\chi$, for which no replica trick is needed.  
Now, if $V[s]$ really vanishes everywhere, then $(4,4,\fourth)_\chi$ is
the correct chiral theory even for nondegenerate masses, and
the  vacuum must therefore remain the standard one.  Thus the assumption of analyticity
includes the assumption that there is no phase change in $(4,``1")_\chi$ as a function
of $s$.

Of course, the assumption of analyticity of $V[s]$ is a nontrivial  one.  It could go wrong in
two ways.  First of all, there may be a connected ``line'' of singularities, an actual ``domain
boundary'' that prevents one from extending $V[s]\!=\!0$ arbitrarily far from $s=0$.  
Of course, \chpt\ or \schpt\ must eventually break down for large enough
quark masses, so it is meaningless to imagine extending \eq{toshow} to mass differences
that put one or more masses outside the range of  chiral perturbation theory.  But here
I am talking about possible singularities that would prevent extending $V[s]\!=\!0$ over the whole
range where \schpt\ applies.  If such a boundary occurred, it would probably imply
a phase change: that the true ground state for  $(4,``1")_{\chi}$ changes discontinuously
from the ground state assumed by $(4,4,\fourth)_\chi$.
Although I cannot rule out this possibility from first principles, it seems rather
unlikely that a phase change would produce small enough discrepancies 
to have escaped detection in the
MILC simulations and their comparison with \schpt\ predictions \cite{MILC}.
But the effects of a phase
change that occurred outside the (rather wide) range of masses or lattice spacings studied by
MILC would probably not have been noticed.  In addition, since the MILC simulations involve
three flavors, the logical possibility exists that a phase change
occurs with four flavors but disappears when the fourth quark is decoupled.
On the positive side,
 note that $(4,4,\fourth)_\chi$ automatically becomes standard
continuum \chpt\ in the continuum limit (see \secref{health}).  Therefore,  
if $V[s]\not=0$ outside some mass region, we must at least have $V[s]\to0$ in the continuum
limit to avoid the  bizarre scenario in which $(4,``1")_{LQCD}$ is a
valid four-flavor QCD theory in some range of quark mass differences but not outside this range.

A second way that the analyticity assumption could go wrong would be the presence
of essential singularities in $V[s]$ for all values of $s$ such that the flavor
symmetry is exact.  For example, one could imagine that $V[s]\propto \exp(-1/V^2_1)$.
Although I cannot rule them out at this point, such singularities seem
implausible to me, since we are expanding around a massive theory in Euclidean space and there are thus
no obvious infrared problems. Note that I am not
assuming that $(4,``1")_\chi$ and $(4,4,\fourth)_\chi$ separately are analytic in $s$
around $s=0$ (or any other degenerate point), only that their difference is. In \secref{conclusions}, I speculate on a 
possible proof of the absence of an essential singularity in $V[s]$ at $s=0$.

The assumption that $V[s]$ is analytic is equivalent to the assumption that
the expansion of $V[s]$ around $s=0$ is convergent. 
The reader may therefore object that this assumption is too strong, since
we do not expect convergent weak coupling expansions in quantum field theories.
It is therefore useful to
review why we believe that usual weak coupling expansions are at best
asymptotic. The main  reason comes from the factorial growth of 
large orders
in perturbation theory \cite{LARGE-ORDER}.  In the current case, however,
the large orders terms in perturbation theory of $V[s]$ in $s$ are not growing factorially ---
in fact they are all zero!  An alternative line of reasoning  for QED is due to
Dyson \cite{DYSON}.  He argued that the expansion in $\alpha$ around $\alpha\!=\!0$
must be asymptotic because $\alpha\!<\!0$ leads to an unstable vacuum and therefore
cannot be smoothly connected to the $\alpha\!>\!0$ region.  In fact, this argument
has been shown to be flawed \cite{BENDER-MILTON}, since it is possible to define
the theory consistently for $\alpha\!<\!0$ and to obtain it by analytic continuation from
$\alpha\!>\!0$.  In any case, however, we have no similar reason to suspect
that the difference of
the chiral theories (or either of the chiral theories itself) becomes
unstable as soon as non-zero mass differences are turned on.

Of course, arguing that we have no reason to expect nonanalyticity in $V[s]$
is far from proving that $V[s]$ is analytic.  This remains an assumption.
Note that it can be turned around: if $V[s]$ is not analytic then, from
\eq{Vs-derivs}, $V[s]\!\not=\!0$, so \schpt\ for
four flavors must be incorrect.  

As mentioned in \secref{notation}, the QCD-level theories  $(4,4,n_R)_{LQCD}$
are used in \secref{expansion} for convenience;
if desired, their use can be eliminated at the expense of an additional
weak assumption about the partially quenched chiral theory.  I now sketch that
argument; 
the example presented in \secref{one-flavor}
can be used as an illustration of this kind of analysis.  
One needs to derive
\eq{chi-derivs-44nR} directly in the chiral theories.
It is not hard to see how to prove this at the chiral level, using a technique
that is basically quark-flow analysis. 
Since the (vector) flavor and replica symmetries are exact in $ (4,4,n_R)_{\chi}$, one
can always follow the replica indices though the \schpt\ diagrams,
starting on one source index and continuing until one reaches another source index
(on the same or a different source).
Each such loop corresponds
exactly to a quark loop at the QCD level and produces
one factor of $n_R$.   The same analysis then needs to be repeated for diagrams with
valence quark indices. 
Note that this
argument assumes that, at the chiral level, mesons made from (fermionic) valence or sea quarks
have identical Feynman rules, except for the counting factors coming from replication.
The ordinary, bosonic, symmetries relating fermionic valence and sea quarks should
guarantee this, as long as such symmetries are not spontaneously broken in the chiral theories.
Since a rigorous
analysis of the partially quenched chiral theory along the lines of Ref.~\cite{GSS} is still lacking,
this absence of symmetry breaking  must be taken as an assumption at this point if one
wants to do without the use of the  QCD-level theories  $(4,4,n_R)_{LQCD}$.
However, it is difficult to see how it could go wrong.

\section{Extension to Fewer than Four Flavors}
\label{sec:extension}

The most interesting cases phenomenologically are three light flavors ($u$, $d$, $s$),
or, at extremely low energies, two light flavors ($u$, $d$).
To extend the above argument to $n_F<4$, we can start by taking one the mass of one of the four quarks
large and using decoupling ideas \cite{APPELQUIST-CARAZZONE}.
Call this quark the charm quark,  with mass $m_c$.
The difficult point here is that the relation \eq{toshow} can only be used
where chiral perturbation theory is applicable, so we cannot just take 
$m_c\to\infty$ on both sides of \eq{toshow} and then appeal to decoupling.

In the real world, we know that the effective coupling of \chpt\ for the strange
quark is roughly $M_K^2/(8 \pi^2 f_\pi^2) \sim 0.2$ \cite{GASSER-LEUTWYLER}, 
with $f_\pi \cong 131\, \MeV$. So it is likely that \chpt\ 
breaks down completely for quark masses that are not very much larger than
the physical strange quark mass, $m_s^{\rm phys}$.
For concreteness, imagine the breakdown occurs at $\sim\!2m_s^{\rm phys}$, in other words for
meson masses greater than $\sim\!700\, \MeV$, which is the mass of a ``kaon'' made
with a strange quark of mass  $2m_s^{\rm phys}$.
I want to decouple the charm quark from \schpt\ before this breakdown occurs,
say at $m_c\!\sim\!1.5m_s^{\rm phys}$.  Since
there is not a lot of room between this value of $m_c$ and $m_s^{\rm phys}$,
it is useful to consider first the case where $m_s$ is significantly
smaller than $m_s^{\rm phys}$.
I try to argue that this $n_F=3$ case
is correctly described by $(3,4,\fourth)_\chi$.

With $m_u$, $m_d$, and $m_s$ all small, I increase $m_c$ to $m_c\sim1.5m_s^{\rm phys}$.
Modulo the assumptions discussed in \secref{results-nf4}, 
the relation $(4,``1")_\chi \doteq (4,4,\fourth)_\chi$ should continue
to hold for $m_c$ in this range.  I then integrate out (decouple) the charm quark degree
of freedom from the chiral theory $(4,4,\fourth)_\chi$. The procedure is completely analogous 
to the way the strange quark is decoupled  from the continuum $SU(3)_L\times SU(3)_R$ chiral
theory to obtain the $SU(2)_L\times SU(2)_R$ theory \cite{GASSER-LEUTWYLER}.
Since this process is perturbative,  there is little doubt
that what  remains after the charm quark is decoupled will be the  $N_f=3$ chiral theory,
$(3,4,\fourth)_\chi$.\footnote{One should also be close enough
to the continuum that the taste-splittings are relatively small,
so that a non-Goldstone meson made from light quarks is significantly
lighter than any meson with a charm quark. This makes the MILC
``coarse'' lattice, with splittings as large as $\sim\!450\,\MeV$ in the chiral limit, 
rather problematic; while the ``fine''
 lattices (largest splittings $\sim\!250\,\MeV$),
should be acceptable.} 
Nevertheless, a check of this assumption in \schpt\ would be reassuring,
and is planned \cite{BERNARD-DU}.

Thus I expect $(4,``1")_{LQCD}$ with $m_c\!\sim\!1.5m_s^{\rm phys}$ to be described at low energy by the chiral
theory $(3,4,\fourth)_\chi$. This should remain true as $m_c$ increases further, say until 
$m_c\!\sim\! 2m_s^{\rm phys}$, which is nominally the largest mass for which 
\eq{toshow} applies.

Consider what happens to $(4,``1")_{LQCD}$ as $m_c$ continues
to increase beyond the applicability of \eq{toshow}. 
When $m_c$ gets to be of order of the cutoff, $m_c\sim 1/a$, one expects that
it will decouple in the usual way from the QCD-level theory, leaving  $(3,``1")_{LQCD}$.
The only effect of the charm quark should be renormalizations of the  $(3,``1")_{LQCD}$
couplings.  The decoupling would be virtually certain if  $(4,``1")_{LQCD}$ were a 
normal theory described by a local lattice action.
Because of the rooting procedure, though, there may be some doubt as to whether decoupling
actually occurs.  We can avoid this concern by increasing $m_c$ still further,
until $m_c \gg 1/a$.  At that point, $m_c$ is much larger than all eigenvalues of
the Dirac operator $D$, and $\root 4 \of {{\rm Det}(D+ m_c)}$ becomes independent
of the gauge field.  Therefore the charm quark certainly decouples
from $(4,``1")_{LQCD}$, leaving $(3,``1")_{LQCD}$.

I am now ready to state the main assumption of this section: {\it As $m_c$ is
increased from  $\sim\!2m_s^{\rm phys}$ to $m_c \gg 1/a$, the low energy physics 
of\/  $(4,``1")_{LQCD}$ is unaffected, except perhaps by renormalizations of the LECs}\/.
Here ``low energy physics'' means the physics of particles with masses and energies
$\ll\! 700\,\MeV$. An alternative way of stating the assumption is to say that
\eq{toshow} continues to be meaningful as $m_c$ is
increased from  $\sim\!2m_s^{\rm phys}$ to $m_c \gg 1/a$, as long as 
$(4,4,\fourth)_\chi$ is interpreted to mean the chiral theory with the
charm quark decoupled, and these theories are only used
at low energy.

I believe the assumption is plausible because the chiral theory shows that $m_c$ is already
decoupled from the low energy physics by $m_c\sim\!1.5m_s^{\rm phys}$.  I
am simply assuming that it stays decoupled as its mass is increased further.

The conclusion then follows immediately:   $(3,4,\fourth)_{\chi}$ is the correct
chiral theory for  $(4,``1")_{LQCD}$ at $m_c\!\sim\!2m_s^{\rm phys}$.  By assumption, it
remains the correct theory as $m_c$ is increased to $\gg 1/a$, at which point
$(4,``1")_{LQCD}$ becomes $(3,``1")_{LQCD}$. Thus
\begin{equation}\eqn{nf3-result}
(3,``1")_\chi \doteq  (3,4,\fourth)_{\chi} \ .
\end{equation}

Note that my decoupling assumption is not only sufficient for \eq{nf3-result}, but also
necessary.  Any new physical effects entering in the region 
 $2m_s^{\rm phys} \ltwid m_c \ltwid 1/a$ are automatically violations of the chiral theory
$(3,4,\fourth)_{\chi}$.

For the moment, \eq{nf3-result} is only true for the three masses $m_u,m_d,m_s \ll m_s^{\rm phys}$, because
these masses needed to be kept small in order to provide a clean decoupling when $m_c \!\sim\! 1.5 m_s^{\rm phys}$.
A line of reasoning parallel to that in \secref{results-nf4} can now be applied: 
Once \eq{nf3-result}
is known to be valid for some range of quark masses, then the difference between
the two theories must vanish everywhere if it is analytic.   The analyticity could be
violated by a phase boundary at some values of the quark mass differences.  However, I
can again point to the MILC simulations \cite{MILC} as evidence against a phase
boundary within the region of parameter space that has been studied.

The arguments (and assumptions) of this section may now be repeated to show 
$(2,``1")_\chi \doteq  (2,4,\fourth)_{\chi}$
and $(1,``1")_\chi \doteq  (1,4,\fourth)_{\chi}$.

\section{Resolution of a Paradox in the One-Flavor Theory}
\label{sec:one-flavor}

An interesting paradox arises from the final result of the previous section for $n_F=1$.
Because of the anomaly, a theory with a single quark flavor should have
no light (pseudo) Goldstone bosons, but only a heavy pseudoscalar, the
$\eta'$.  On the other hand, the \schpt\ for a single rooted staggered flavor
contains 16 pseudoscalars (``pions"), of which only one, the taste-singlet, is
heavy.  The weightings of the contributions of the different pions in this rooted theory
have factors of $1/4$ 
compared to those in the unrooted, four-taste theory,
but all the pions certainly 
contribute in both rooted and unrooted cases at finite lattice spacing. 
Even in pure-glue correlation
functions, the light pions will
appear as intermediate states.\footnote{I thank Andreas
Kronfeld for emphasizing to me the importance of addressing this paradox.}

If my previous arguments are correct, then we know the chiral theory for
a single flavor of rooted staggered quarks, and it will produce the correct
chiral theory in the continuum limit for QCD with a single flavor.  The only
way out of the paradox is therefore that  the light pions decouple
from physical correlation functions in the continuum limit.
In this section, I present a particular example to show
in detail how the decoupling takes place at leading order in the chiral theory.  
This is a special case of the calculations of scalar, isoscalar
correlators for various numbers of flavors worked out in Ref.~\cite{SCALAR}, and will be discussed in more
detail there.

Gluon or glueball interpolating fields at physical momenta ($\ll1/a$) couple
only to taste-invariant combinations of the quark fields.
To mock up a pure-glue correlation function, we add a taste-singlet scalar 
source to the rooted one-flavor theory:
\begin{equation}\eqn{one-flavor-source}
	\cL_{\rm source} =  
s(z) \bar \Psi(z)  \Psi(z)\ .
\end{equation}
Here $ s(z)$ is basically the same as the sources  considered previously in \eq{source-41},
except that there are no flavor indices since $n_F=1$.

The generating functional of this theory, $(1,``1")_{LQCD}$, is obtained by computing the
fermion determinant in the presence of the source $ s$, taking its fourth root,
and then integrating over gauge fields.   In order to show explicitly the factors 
resulting from the rooting, I will take the $R^{th}$ power of the determinant, and only
set $R=1/4$ at the end.  The generating functional is thus given by:
\begin{equation}\eqn{generating-11}
(1,``1")_{LQCD}= \frac{\int \cD\! A\, \exp\{-S_G(A) + R\, \tr\left( \ln\,(D + m +  s)\right)\}}
{\int \cD\! A\, \exp\{-S_G(A) + R\, \tr\left( \ln\,(D + m )\right)\}} \ ,
\end{equation}
where $D$ is the Dirac operator for the staggered field, $m$ is its mass, $A$ represents
the gauge fields, with action $S_G(A)$, 
and $\cD\! A$ is the gauge measure.  As usual, one should imagine
that additional  valence quark fields (and the corresponding commuting ghost quark fields to
cancel the valence determinant \cite{QUENCH}) are included as needed.
	
Note that $R$ in \eq{generating-11} is a parameter appearing in the QCD-level generating function.  It is
logically independent from $n_R$, which is the number of sea-quark replicas and is introduced (later)
as a way of representing the rooting trick at the chiral level.  
Of course, in the end we want
to set both $R$ and $n_R$ to $1/4$.

I want to calculate
\begin{equation}\eqn{Gxy}
G(x\!-\!y) = \left(\frac{\partial}{\partial  s(x)}
\frac{\partial}{\partial  s(y)}\; (1,``1")_{LQCD}\right)_{\hspace{-0.15cm} s=0} 
\hspace{-0.15cm}-\left(\frac{\partial}
{\partial  s(x)}\; (1,``1")_{LQCD}\right)_{\hspace{-0.15cm} s=0}
\hspace{-0.1cm}\left(\frac{\partial}
{\partial  s(y)}\; (1,``1")_{LQCD}\right)_{\hspace{-0.15cm} s=0}\hspace{-0.25cm}.\phantom{\Bigg\vert_p}
\end{equation}
The second term subtracts off the limit 
at infinite separation,
proportional to $\langle\bar\Psi\Psi\rangle^2$.
What remains is just the connected part of the correlation function at the QCD level.
We are interested in the lightest particles that appear as intermediate states
in the decay of $G(x\!-\!y)$ at large $|x\!-\!y|$.

There are two possible contractions contributing to the first term in
$G(x\!-\!y)$, as in \eq{example-ii}; while there is only one contraction in each of
the factors in the second term.
Introducing valence quarks $q_\alpha$, degenerate with the sea quarks,
and corresponding valence sources $ \sigma^{\alpha\beta}$,
I rewrite the contributions in terms of valence quark contractions. With 
$\alpha\not=\beta$, one has
\begin{eqnarray}
G(x\!-\!y)&=& R\left(\frac{\partial}{\partial \sigma^{\alpha\beta}(x)}
\frac{\partial}{\partial \sigma^{\beta\alpha}(y)} 
(1,``1")_{LQCD}\right)_{\hspace{-.15cm}\sigma=0} 
+R^2\left(\frac{\partial}{\partial \sigma^{\alpha\alpha}(x)}\frac{\partial}{\partial \sigma^{\beta\beta}(y)}
(1,``1")_{LQCD}\right)_{\hspace{-.15cm}\sigma=0} \nonumber \\
&&\hspace{1cm}
-\;
R^2\left(\frac{\partial}{\partial \sigma^{\alpha\alpha}(x)} (1,``1")_{LQCD}\right)_{\hspace{-.15cm}\sigma=0} 
\left(\frac{\partial}{\partial \sigma^{\beta\beta}(y)}
(1,``1")_{LQCD}\right)_{\hspace{-.15cm}\sigma=0} 
\eqn{Gxy-valence}
\end{eqnarray}
Here and below the sea quark source $ s$ has been set to zero.

\begin{figure}[t]
\resizebox{6.0in}{!}{\includegraphics{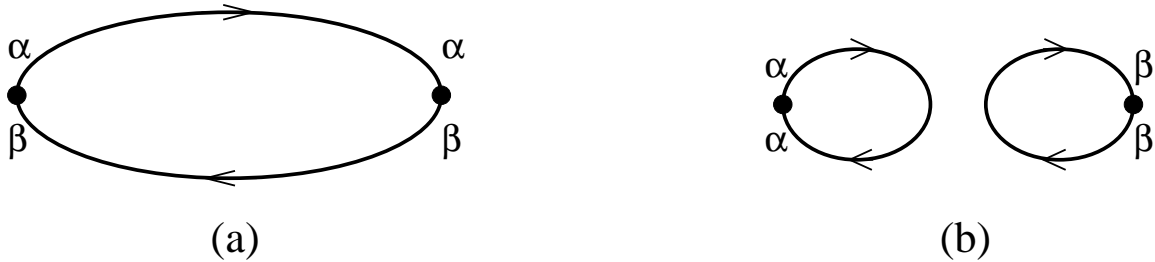}}
\caption{Valence quark contractions in the scalar propagator $G(x-y)$, corresponding
to \protect{\eq{Gxy-valence}}.  The solid dots represent the source $ \sigma$.
Only the valence quark lines are shown; completely disconnected contributions
to (b) should be omitted.
\label{fig:contractions}}
\end{figure}

The contractions in \eq{Gxy-valence}
are shown in \figref{contractions}. The first term (multiplied by $R$) is
pictured in \figref{contractions}(a); the second term (multiplied by $R^2$), in \figref{contractions}(b).
Arbitrary numbers of gluon propagators and sea quark loops are implied, except that the
third term  in \eq{Gxy-valence}
is taken into account by omitting disconnected contributions to
\figref{contractions}(b).  

By the arguments of this paper, we should be able to calculate the
low-mass contributions to $G(x\!-\!y)$
using the appropriate \schpt.  That theory is $(1,4,n_R)_\chi$,
with $n_R$ set to $1/4$ after the calculation to implement the replica
trick.  
 I append to $(1,4,n_R)_\chi$ the valence degrees of freedom
associated with the two flavors  $\alpha$ and $\beta$ in \eq{Gxy-valence}, as well as the
corresponding two commuting ghost quarks.
Including taste degrees of freedom, the symmetry group  of $(1,4,n_R)_\chi$ is then
$SU(4n_R+8|8)_L \times SU(4n_R+8|8)_R$.  
Following the notation of \cite{AUBIN-BERNARD}, I define
a meson field $\Phi$, which is a $(4n_R+16) \times (4n_R+16)$ Hermitian matrix,
and the unitary  matrix $\Sigma=\exp(i\Phi/f)$, where $f$ is the LO pion decay constant.
With $a$ and $b$ flavor indices, running over both valence and sea flavor,
we may write
\begin{equation}\eqn{phi-def}
\Phi_{ab} = \sum^{16}_{\Xi=1} \Phi_{ab}^\Xi\; t_\Xi\;,
\end{equation}
where the $ \Phi_{ab}^\Xi$ correspond to mesons of specific taste and flavor, and
$t_\Xi$ are the 16 taste generators
\begin{equation}\eqn{taste-generators}
\{t_\Xi\} = \{I,\;\xi_\mu,\; \xi_{\mu\nu}(\mu>\nu),\; \xi_{\mu}\xi_5,\; \xi_5\} \ .
\end{equation}
with $\xi_\mu$ the $4\!\times\!4$ taste matrices that correspond to the Dirac gamma matrices.
All quarks (sea and valence) are degenerate,
with mass $m$.

$G(x\!-\!y)$ will be calculated at leading order (LO) in \schpt.  
The valence source $ \sigma$ couples exactly like the valence mass term, giving a   
contribution to the LO Euclidean chiral Lagrangian:
\begin{equation}\eqn{chiral-source-term}
\cL_{\rm source} = -\frac{1}{4}\mu f^2\;  \sigma_{\tau\rho}\;\tr\!
\left(\Sigma_{\rho\tau}+\Sigma^\dagger_{\rho\tau}\right)\ ,
\end{equation}
where $\mu$ is the chiral condensate,
$\rho$ and $\tau$ are valence flavor indices (summed over valence-quark, but not ghost-quark,
flavors), and $\tr$ indicates a trace over taste indices only. There are also terms
quadratic in $\sigma$ appearing in the next-to-leading order Lagrangian; they may be ignored
because they contribute only to contact terms in $G(x-y)$ to the order we are working.

To convert \eq{Gxy-valence} to the chiral level, we just replace
$(1,``1")_{LQCD}$ with  $(1,4,n_R)_\chi$. Then, using \eq{chiral-source-term}, 
and expanding $\Sigma$ and $\Sigma^\dagger$ to second order in $\Phi$, we have
\begin{equation}\eqn{Gxy-chiral}
G(x\!-\!y) = R\mu^2 \Big\langle \Phi^\Xi_{\alpha a}(x)\,\Phi^\Xi_{a\beta}(x)\; \Phi^{\Xi'}_{\beta b}(y) 
\,\Phi^{\Xi'}_{b\alpha}(y)\Big\rangle
+ R^2\mu^2 \Big\langle \Phi^\Xi_{\alpha a}(x)\,\Phi^\Xi_{a\alpha}(x)\; \Phi^{\Xi'}_{\beta b}(y)\, 
\Phi^{\Xi'}_{b\beta}(y)\Big\rangle_{\rm conn}\ ,
\end{equation}
where there are implicit sums over the taste indices $\Xi$ and $\Xi'$, as well as 
over the (generic) flavor indices $a$ and $b$,
but not over the valence flavor indices $\alpha\not=\beta$.  The subscript ``conn'' on the second
term means that only those {\it meson}\/ contractions that connect the source points $x$ and $y$ should be 
included.
This restriction arises from the cancellations due to the last term in \eq{Gxy-valence}.
(The first term \eq{Gxy-chiral} does not need a ``conn'' subscript because the valence indices
require that all contractions connect $x$ to $y$.)  Cancellations of the disconnected pieces are also
responsible for the absence in \eq{Gxy-chiral} of contributions from the ``1'' terms in
the expansion of $\Sigma+\Sigma^\dagger$.

\begin{figure}[t]
\resizebox{6.0in}{!}{\includegraphics{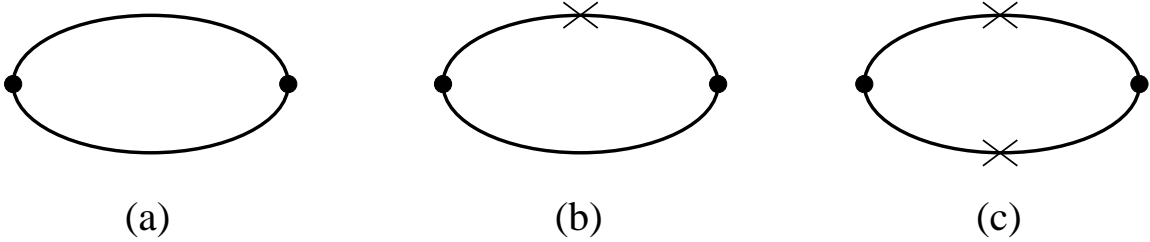}}
\caption{Lowest order \schpt\ meson diagrams coming from
\protect{\eq{Gxy-chiral}}, and corresponding to \protect{\figref{contractions}}.
As in \protect{\figref{contractions}}, a solid dot is  a source, $ \sigma$.
The cross represents one or more insertions of a ``hairpin'' vertex, and hence indicates
a meson propagator that is disconnected as a quark-flow diagram.  
\label{fig:mesons}}
\end{figure}

\Figref{mesons} shows the LO (one-loop) meson diagrams contributing to \eq{Gxy-chiral}.
The crosses indicate a
presence of one or more  ``hairpin'' vertices, which can appear on
flavor-neutral meson lines.  In the quark-flow sense, propagators
without hairpin vertices are connected; while those with at least one
hairpin are disconnected. (See for example Fig.~1 in the first reference in \cite{AUBIN-BERNARD}.)
Note however that even a hairpin diagram is connected in the QCD (or meson) sense, since
gluons connect the two quark lines.

In \schpt, hairpin vertices are of two types:
The first is due to the anomaly and affects only taste-singlet meson propagators. 
In the notation of Ref.~\cite{AUBIN-BERNARD}, it has
strength $4m_0^2/3$ for arbitrary numbers of flavors. The anomaly contribution
to the mass-squared of the $\eta'$ is proportional
to $m^2_0$, with the proportionality constant depending on the total number of flavors (more
precisely in this case, on the number of replicas $n_R$).
The second kind of hairpin is due to the taste-violating operators that appear in \schpt.
These hairpins affect only taste-vector and taste-axial-vector mesons at LO, and
have strength proportional to $a^2$.  Due to the explicit factors of $a^2$, the contributions
of such taste-violating hairpins automatically vanish in the continuum limit. Since I 
am interested in the restoration of physical unitarity in the
continuum, I omit the taste-violating hairpins here; they of
course are included in a complete LO calculation \cite{SCALAR}.

We now go to momentum space. 
The (quark-flow) connected propagator carrying Euclidean momentum $p$ is
\begin{equation}\label{eq:connected-prop}
        \left\langle\Phi^{\Xi}_{ab}(-p)\;\Phi^{\Xi'}_{b'a'}(p)\right\rangle_{\hspace{-.05cm}\rm conn} =
        \frac{\delta_{a,a'}\,\delta_{b,b'}\, \delta_{\Xi,\Xi'}}{p^2 + M^2_\Xi } \ ,
\end{equation}
where $M_\Xi$ is
the tree-level mass of a taste-$\Xi$ meson 
\begin{equation}\label{eq:pi-masses}
       M_\Xi^2  = 2\mu m  + a^2\Delta_\Xi\ ,
\end{equation}
with $\Delta_\Xi$ the taste splitting. All quarks are degenerate so there
is no need to specify the flavor in \eq{pi-masses}.

Keeping only the taste-singlet disconnected meson propagator,  we have
\begin{equation}\label{eq:disconnected-prop}
        \left\langle\Phi^{\Xi}_{ab}(-p)\;\Phi^{\Xi'}_{b'a'}(p)\right\rangle_{\hspace{-.05cm}\rm disc}=
        \delta_{a,b}\,\delta_{b',a'}\, \delta_{\Xi, I}\,\delta_{\Xi',I}\, \cD^I(p) \ ,
\end{equation}
where \cite{AUBIN-BERNARD}
 \begin{equation}\label{eq:Disc}
\cD^I(p) = -\frac{4m_0^2}{3}\, \frac{1}{(p^2+M_I^2)}\,
 \frac{1}{(p^2+M^2_{\eta'_I})}\ ,
\end{equation}
with 
\begin{equation}\eqn{metap}
M^2_{\eta'_I} = M^2_I + n_R \frac{4m_0^2}{3} \ .
\end{equation}
Note that, by definition,
$M_I$ is the mass of any taste-singlet meson before including the
effect of the anomaly hairpin.  Thus all 16 of the masses listed in \eq{pi-masses}, including
$M_I$, become equal in the continuum limit.
 The $\eta'_I$,  on the other hand, is the one meson that is a flavor (more precisely, replica) singlet
as well as a taste singlet. Its mass $M_{\eta'_I}$ can be found either by diagonalizing
the complete LO mass matrix including the anomaly term, or by 
summing the geometric series of hairpin interactions.

One could take the limit $m_0^2\to\infty$ in \eq{Disc} to decouple the $\eta'$,
since it is at least as heavy as particles we have integrated
out of the chiral theory (\eg the $\rho$).  However, I prefer to leave $m_0$
finite so we may see explicitly how the $\eta'$ remains after all the light mesons cancel in
the continuum limit.

\begin{figure}[t]
\resizebox{5.0in}{!}{\includegraphics{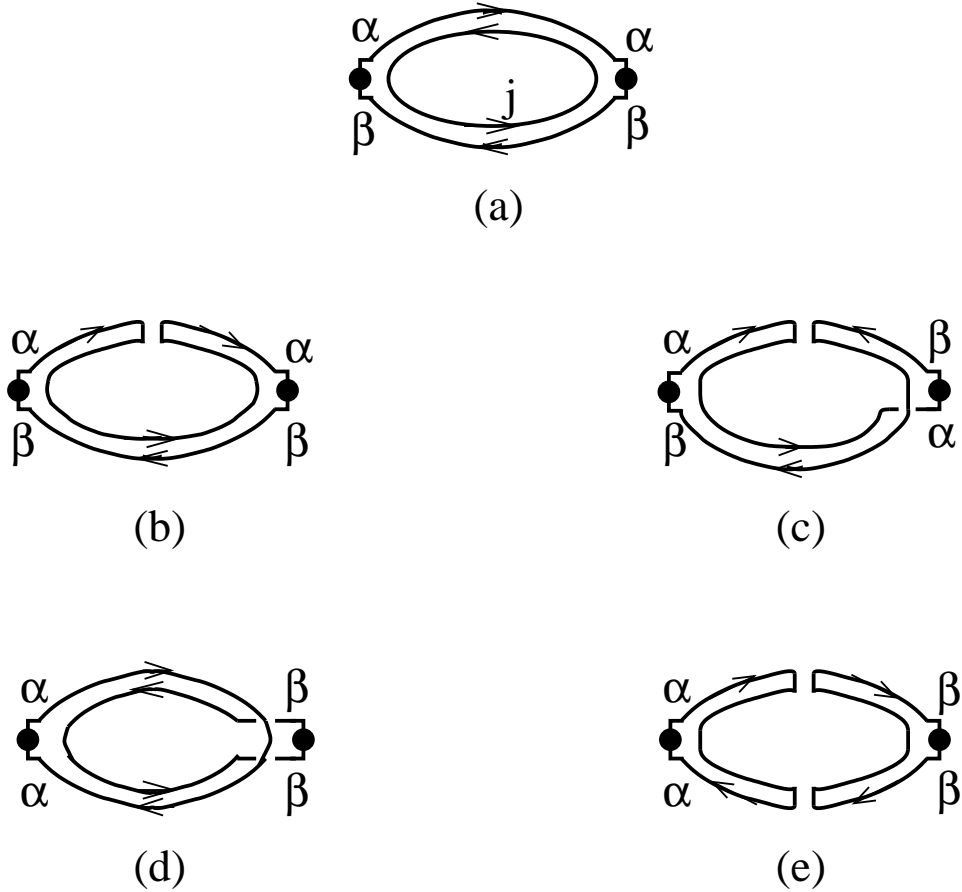}}
\caption{Quark flow diagrams corresponding to the  \schpt\ contributions of \protect{\figref{mesons}}.
Not shown are two additional diagrams that are very similar to (b) and (c) but have the
roles of valence quarks $\alpha$ and $\beta$ interchanged.
Diagrams (a) and (d) have no hairpin vertices and correspond to \protect{\figref{mesons}}(a);
diagrams (b) and (c) have one hairpin vertex and correspond to \protect{\figref{mesons}}(b);
while diagram (e), with two hairpin vertices, corresponds to \protect{\figref{mesons}}(c).
In meson lines with hairpin vertices, a summation of sea-quark loop insertions is implied.
\label{fig:quark-flow}}
\end{figure}

I now consider the meson contractions that contribute to \eq{Gxy-chiral}.
Although it is not necessary to use a quark-flow
picture here, since adjustment for the rooting is automatically taken into
account by setting $n_R=1/4$, quark flow  gives a nice physical picture. In \figref{quark-flow}, 
I therefore show the
quark-flow diagrams that correspond to various meson contractions.  
\Figref{quark-flow}(a), (b), and (c)
come from the $R$ term in \eq{Gxy-chiral}.  Note that, like 
\figref{contractions}(a) from which they arise,
\figref{quark-flow}(a), (b), and (c) have a single valence-quark loop connecting the sources
(shown by solid dots). Similarly, \figref{quark-flow}(d) and (e) come  from the $R^2$ term
in \eq{Gxy-chiral}, and, like  
\figref{contractions}(b), have two separate valence-quark loops.

When $a$ and $b$ in \eq{Gxy-chiral} are sea quark flavors
$i$ and $j$,\footnote{I use Latin indices from the middle of the alphabet ($i$, $j$, \dots) for
sea quark flavors (replicas, here), Greek indices ($\alpha$, $\beta$, \dots) for fermionic valence quark flavors, and Latin indices
from the beginning of the alphabet ($a$, $b$, \dots) for generic valence or sea flavors.} connected meson
propagators are only possible for the term proportional to $R$ in  \eq{Gxy-chiral}, and require
$i=j$.
(In the $R^2$ term the flavors do not match up.)  This generates \figref{quark-flow}(a), which
is proportional to $n_R$ due to the sum over sea quark flavors $j$.

When $a$ and $b$ are valence quark indices, connected contractions like those in  \figref{quark-flow}(a) are also
possible, but there is a cancellation between valence quarks and ghost quarks, as follows in the
quark-flow picture from
the fact that  \figref{quark-flow}(a) has a virtual loop.  

One additional contraction with only connected
meson propagators comes from the $R^2$ term in 
\eq{Gxy-valence} when $a=\beta$ and $b=\alpha$.  In the quark flow picture, this gives diagram
 \figref{quark-flow}(d), which is constructed entirely from valence quarks and therefore generates
no factors of $n_R$.

Contractions with a single disconnected meson
propagator are generated only by the $R$ term in \eq{Gxy-chiral}.
It gives diagram  \figref{quark-flow}(b) or
 the $\alpha\leftrightarrow\beta$ version
when $a=b=\alpha$ or $a=b=\beta$, respectively. Similarly, it gives diagram  
\figref{quark-flow}(c) or the $\alpha\leftrightarrow\beta$ version
when $a=\alpha$, $b=\beta$ or $a=\beta$, $b=\alpha$, respectively. These four terms, which correspond
at the meson level to \figref{mesons}(b), can be seen to have the same numerical value after
adjusting the loop momentum assignment.

Finally, the $R^2$ term in \eq{Gxy-chiral} gives diagram \figref{quark-flow}(e) 
when $a=\alpha$ and $b=\beta$. There is an overall symmetry factor of $2$  in this case.

We can now add the various contributions to $\tilde G(q)$, the Fourier transform
of $G(x-y)$, resulting in:
\begin{eqnarray}
\tilde G(q) &=& \mu^2 \int \frac{d^4 p}{(2\pi)^4}\;\left\{\vbox to 24pt{}\left(Rn_R + R^2\right)
\sum_\Xi \frac{1}{\left(p^2+M^2_\Xi\right)\left(\left(p+q\right)^2+M^2_\Xi\right)} \right.
\nonumber \\
&&-\; \frac{2R\left(4m_0^2/3\right)}{\left(p^2+M^2_I\right)\left(\left(p+q\right)^2+M^2_I\right)}
\left(\frac{1}{p^2+M^2_{\eta'_I}}
+ \frac{1}{\left(p+q\right)^2+M^2_{\eta'_I}}\right) \nonumber \\
&&+\; \left.\vbox to 24pt{}\frac{2R^2\left(4m_0^2/3\right)^2}{\left(p^2+M^2_I\right)\left(\left(p+q\right)^2+M^2_I\right)
\left(p^2+M^2_{\eta'_I}\right)\left(\left(p+q\right)^2+M^2_{\eta'_I}\right)}\; \right\}\ ,
\eqn{Gq-raw}
\end{eqnarray}
The first line in \eq{Gq-raw} comes from \figref{quark-flow}(a) and (d), which give the
$Rn_R$ and the $R^2$ terms, respectively; the second line, from \figref{quark-flow}(b), (c),
and their $\alpha\leftrightarrow\beta$  versions; the last line, from \figref{quark-flow}(e).
Note that the negative sign of the anomaly hairpin,
\eq{Disc}, makes the second
line negative and leads to the possibility of cancellations among the various light pions.
It is not an accident that the hairpin is negative: It is required in order to
give a positive mass to the $\eta'$ when the geometric series of
insertions is summed. 

Using, from \eq{metap},
\begin{equation}\eqn{m0-rewrite}
\frac{4m_0^2}{3} 
= \frac{1}{n_R}\left[ \left(p^2 + M^2_{\eta'_I}\right) -  \left(p^2 + M^2_I\right)\right] 
= \frac{1}{n_R}\left[ \left(\left(p+q\right)^2 + M^2_{\eta'_I}\right) -  \left(\left(p+q\right)^2 + M^2_I\right)\right], 
\end{equation}
one can rewrite \eq{Gq-raw} in a form that shows more clearly how the continuum limit works:
\begin{eqnarray}
\tilde G(q) &=& \mu^2 \int \frac{d^4 p}{(2\pi)^4}\;\left\{\vbox to 24pt{}
\frac{2R^2}{n_R^2}
\frac{1}{\left(p^2+M^2_{\eta'_I}\right)\left(\left(p+q\right)^2+M^2_{\eta'_I}\right)}
\right.  + \nonumber \\
&&\hspace{-1.5cm}+\left(Rn_R + R^2\right)
\sum_\Xi \frac{1}{\left(p^2+M^2_\Xi\right)\left(\left(p+q\right)^2+M^2_\Xi\right)} 
-\left(\frac{4R}{n_R}-\frac{2R^2}{n_R^2}\right)
	\frac{1}{\left(p^2+M^2_I\right)\left(\left(p+q\right)^2+M^2_I\right)}\nonumber \\
&&\hspace{-1.5cm}+\left.\vbox to 24pt{}
\left(\frac{2R}{n_R}-\frac{2R^2}{n_R^2}\right)
   \left(\frac{1}{\left(p^2+M^2_I\right)\left(\left(p+q\right)^2+M^2_{\eta'_I}\right)}
   +\frac{1}{\left(p^2+M^2_{\eta'_I}\right)\left(\left(p+q\right)^2+M^2_I\right)} \right)
\right\}.
\eqn{Gq}
\end{eqnarray}
Setting $R=1/4=n_R$, the last line of \eq{Gq} vanishes immediately.  The light pions
then contribute only in the second line.  In the continuum limit all 16 of the
light masses $M_\Xi$ become degenerate, and the two terms on the second line
cancel also.  The remainder, the first line, comes from the exchange of two 
heavy singlet mesons ($\eta'_I$), and indeed has the same normalization as would be found
for this correlation function using a continuum one-flavor chiral theory.
This resolves the apparent one-flavor paradox, showing that it does not
provide a counterexample to the arguments of this paper.

\section{Consequences}
\label{sec:consequences}

\subsection{Health of the Rooted Theory}
\label{sec:health}

With the usual assumption that
taste symmetry is restored in the continuum limit for unrooted staggered
quarks,  $(n_F,4,n_R)_{\chi}$ becomes ordinary chiral perturbation theory
for $4n_F\cdot n_R$ ``flavors'' in the continuum limit.  This follows immediately
from the fact that, for a given combination of quark flavors, all 16 taste
pions become degenerate in the continuum limit (before the effects of the
anomaly are included, which affects only the taste singlet, flavor singlet
meson, as always).  Then taking $n_R\to\fourth$ order by order
necessarily produces standard, continuum \chpt\ for $n_F$ flavors.
All existing \schpt\ calculations \cite{LEE-SHARPE,AUBIN-BERNARD,PRELOVSEK,HEAVYLIGHT,SCHPT-OTHER,SCHPT-BARYONS}
have this expected continuum limit.

The statement that  $(n_F,4,\fourth)_{\chi}$ is the correct chiral theory for
 $(n_F,``1")_{LQCD}$ (for $n_F\le4$) therefore has important implications for the validity of the
rooting procedure itself.  
Since \schpt\ becomes standard \chpt\ in the continuum limit, 
the low energy
(light pseudoscalar meson) sector of $n_F$-flavor lattice  QCD with rooted staggered quarks is, in the
continuum limit, indistinguishable in its structure from
that of ordinary $n_F$-flavor QCD.   There are no violations of unitarity, and
no introduction of unphysical nonlocal scales.

Of course, the chiral perturbation theory arguments presented in this paper
do not address possible sickness due to rooting that would appear in
sectors of the theory not described by \chpt.  Nevertheless, the extension
of my arguments to at least some sectors other than that of the light pseudoscalar mesons
seems possible.  In particular, heavy-light physics can be described
by the addition of a {\it valence}\/ heavy quark with a nonstaggered action to the
existing \schpt\ framework \cite{HEAVYLIGHT}.  As such, the arguments in \secref{details}
should also apply in the heavy-light case with $n_F=4$ sea quarks, implying that it too is 
free from unphysical effects in
the continuum limit.  Further, I see no obvious problems
with an extension to $n_F<4$,
since the heavy-quark can be treated by heavy quark effective theory at both the QCD and the chiral level,
and thus does not introduce a new scale that would interfere with decoupling.
The case of baryons, described by staggered heavy-baryon
chiral perturbation theory \cite{SCHPT-BARYONS}, also
seems straightforward for $n_F=4$. However, the artifice of increasing the number of
colors at the QCD level is not applicable in this case, because it changes the nature
of the baryons.  Therefore, any counting arguments like those in \secref{expansion} would need
to be performed at the chiral level only. In addition, it is not obvious that one can use
decoupling to analyze the $n_F<4$ cases, 
since we would now have the baryon mass scale at the QCD level between 
$700\,\MeV$ and $1/a$.

One {\it caveat} should be added to the discussion of this section:
Since the chiral expansion expresses physical quantities 
in terms of unknown LECs, the statement that \schpt\ is valid
does not by itself$\,$ imply that the  LECs generated by the rooted staggered
theory take on their correct (real QCD) values in the continuum limit.  
On the other hand, in the four-flavor case we do know that the LECs are 
correct in the degenerate
case. This follows from universality since the degenerate action is local.
But the LECs  are by definition mass independent, so if  \schpt\ is
indeed the right chiral theory for four nondegenerate flavors, the
LECs are {\it per force}\/ also correct.   With fewer than four flavors, though,
my assumptions on decoupling do not appear to be strong enough to continue
to guarantee correct LECs.  For that one 
would need to show universality at the lattice  QCD level (see for example Ref.~\cite{SHAMIR}), 
or to argue from the agreement of rooted staggered simulations with experiment \cite{PRL}.
Of course, numerical checks against experiment are not proofs, and they
run the risk, at least in principle, of confounding small violations of 
universality due to rooted staggered quarks with small violations of the Standard Model.  
Such checks will become more convincing when one can see agreement 
between at least two different lattice fermion  approaches.

\subsection{A ``Mixed'' Theory?}
\label{sec:mixed}

In current dynamical staggered simulations \cite{MILC}, the fourth-root trick
is applied to the sea quarks, while the valence quarks are described by ordinary staggered
fields. In this section, I call this situation a ``rooted-staggered theory''
for simplicity. Because valence and sea quarks are treated differently,
it has been suggested \cite{KENNEDY} that rooted-staggered theories
fall into the class called ``mixed,'' where the valence and sea quarks
have fundamentally different lattice actions.  
In mixed theories the mass
renormalizations of sea and valence quarks are different, meaning in particular
that there is no simple way to ensure that sea and valence quarks
have the same physical mass.  Further, the continuum
symmetries that would rotate valence and sea quark into each other are violated
by discretization effects. This implies, for example,
that even if the quark masses are adjusted to make the mass of
a meson with two valence quarks equal to the mass of a meson with two sea quarks, the mass of
a meson with one valence and one sea quark will be different by
$\cO(a)$ or $\cO(a^2)$ terms.  Such terms show up 
in new operators in the  \chpt\ for the mixed theory \cite{MIXED}.

I claim, however, that the rooted-staggered case 
is not a mixed case, but in fact resembles much more closely  a partially
quenched theory, where the symmetries between valence and sea quarks are violated
only by explicit differences in quark masses.  

First of all, I sketch a proof that the renormalization
of sea and valence quark masses are the same to all orders in (weak-coupling) perturbation theory.
Imagine we have determined the mass counterterm for a valence quark up to an including a given order
in perturbation theory.  I need to show that the same mass counterterm will
work to renormalize the mass on a sea quark line that appears as a loop inside some
other diagram.
Go inside the diagram, and
draw a box around a self-energy insertion on a sea quark line. 
As remarked in \secref{replica}, the replica trick shows that the rooting procedure 
simply multiplies each sea quark loop by $1/4$ in  perturbation theory, so the self-energy insertion
as well as the associated mass counterterms on that line are all multiplied
by the same overall factor of $1/4$, compared to the  corresponding self-energy insertion
and mass counterterms on a valence line.  
Thus the same counterterms work in both cases.
Of course there may be additional  factors of 1/4 for any sea quark loops that appear in
sub-sub-diagrams. But these  will be the same for a sea quark line as for a valence
line.

The argument in the proceeding paragraph is based on weak-coupling perturbation theory.
Could there be ``mixed effects'' that show up only nonperturbatively?  I cannot
answer that question for general nonperturbative effects, but I can answer it 
--- modulo the assumptions
in \secrefs{details}{extension} ---
for the large
class of effects described by the chiral theories.
The appropriate chiral
theory  is $(n_F,4,\fourth)_\chi$, which is obtained order by order from
$(n_F,4,n_R)_\chi$.  The latter theories have symmetries
interchanging
valence and sea quarks. For $n_V$ flavors of valence staggered quarks,
the full symmetry group is in fact 
$SU(4n_Rn_F \hspace{-.05cm}+\hspace{-.05cm} 4n_V|4n_V)_L \times SU(4n_Rn_F \hspace{-.05cm}+\hspace{-.05cm} 4n_V|4n_V)_R$.  The taste
symmetries are broken on the lattice at $\cO(a^2)$, but the ``flavor subgroup''\footnote{This flavor
subgroup is described in the  first paper in Ref.~\cite{AUBIN-BERNARD}, but is there
called the ``residual chiral group.''  It has been generalized here to take into account
the partially quenched context.}
$U(n_Rn_F \hspace{-.05cm}+\hspace{-.05cm} n_V|n_V)_\ell \times U(n_Rn_F \hspace{-.05cm}+\hspace{-.05cm} n_V|n_V)_r$ is exact
up to quark mass terms.
Extra chiral operators that would split valence-sea mesons from sea-sea or 
valence-valence mesons are forbidden by these symmetries.
Since such operators are absent for all $n_R$, they can have no effect when we take $n_R\to\fourth$.
In particular, corresponding sea-sea, valence-valence, and valence-sea mesons are all degenerate
(when the quark masses are degenerate) in  $(n_F,4,n_R)_\chi$, and therefore in $(n_F,4,\fourth)_\chi$.
Thus, at least within the context of chiral perturbation theory, the rooted-staggered theory
behaves like a partially quenched theory, {\it not}\/ like a mixed theory.

One does have to be careful in defining the word ``corresponding'' in the previous
paragraph.  The valence sector of a rooted-staggered theory is ``richer'' than the sea sector, in that it
includes particles in the continuum limit whose sea-sector analogues have decoupled
from the physical theory.
This is not surprising, since the purpose of the fourth root is to reduce four
sea quark tastes to one, and there is no fourth root taken in the valence sector.
A simple example of this behavior can be seen from the result in \secref{one-flavor}.
If one adds together the valence contractions in \eq{Gxy-valence} without the
extra factor of $R$ relating the last two terms to the first, then one gets a 
valence Green's function with no sea-quark analogue.  
Intermediate light (pseudo-Goldstone)
mesons will appear as intermediate states of this Green's function in the continuum limit.  
In this sense, the rooted-staggered theory, 
is inherently ``partially quenched,'' even in limit of equal valence
and sea masses.  In a normal partially quenched theory, one can always take more
valence quarks than there are sea quarks, so one has the option of creating valence
states that have no analogue in the sea-quark sector.   The main difference here is that
one has no choice in the matter:
The physical sea-quark subspace is always a proper subspace of the complete theory in the 
continuum limit. 

\section{Conclusions and Discussion}
\label{sec:conclusions}

Under certain assumptions that I repeat below, I have shown in this paper that staggered
chiral perturbation theory (\schpt) correctly describes the low energy physics of four 
or fewer flavors of rooted staggered quarks.  The \schpt\ theory $(n_F,4,\fourth)_\chi$
takes into account the fourth-root procedure by the replica trick (or equivalently, 
by quark-flow analysis).  At finite lattice spacing, \schpt\ reproduces unphysical
features of the rooting that may perhaps best be described as violations of unitarity,
with unwanted intermediate states contributing to amplitudes.  This is clearly
seen in Ref.~\cite{PRELOVSEK} or in the example presented in \secref{one-flavor}
\cite{SCALAR}.  

Because \schpt\ becomes standard \chpt\ in the continuum limit, 
the unitarity violations seen in \schpt\ at nonzero $a$ must go away when $a\to0$.
If \schpt\ is indeed the correct chiral theory for rooted staggered quarks, then
this implies that 
the low energy (pseudoscalar meson) sector of lattice  QCD with rooted staggered quarks is, in the
continuum limit, indistinguishable in its structure from
that of ordinary QCD.   There are no violations of unitarity, and
no introduction of unphysical nonlocal scales.
This would not, by itself, show that the rooting procedure is valid, because
there could be problems in sectors of the theory not described by
chiral perturbation theory.  Nevertheless, it would significantly reduce the possible ways 
in which rooted staggered quarks could go wrong. 

My \schpt\ results also give support to the statement that
the theory with staggered valence quarks and rooted staggered sea quarks is not
a ``mixed'' theory.  Like a partially quenched theory with the same action
for the valence and sea quarks, the rooted staggered theory has flavor symmetries
rotating sea and valence quarks into each other.  These symmetries may be
broken in the usual way by mass terms, but they are not broken by lattice corrections.

The starting point of my argument was the observation that four flavors of 
degenerate staggered quarks simply reduce to a single flavor when the
fourth root of the determinant is taken.  To make use of this observation,
I needed several assumptions, the most important of which are:

\begin{itemize}
\item[1)\ ]{} The taste symmetry is restored in the continuum limit of the
normal, unrooted staggered theory.  

\item[2)\ ]{} The difference $V[s]$ between the \schpt\ theory for four flavors $(4,4,\fourth)_\chi$
and the  true chiral theory  $(4,``1")_\chi$ is analytic in $s$ (for space-time
independent $s$), up to possible isolated
singularities. 

\item[3)\ ]{} As a single quark mass  (``charm'') is
increased beyond the point at which it has decoupled from the chiral theory
to a scale much larger than the
lattice cutoff, the low energy physics
of  $(4,``1")_{LQCD}$ is unaffected, except perhaps by renormalizations of the LECs.
\end{itemize}

I consider assumption 1) to be noncontroversial, and there is a lot
of numerical evidence for it, but it has not been rigorously proven.
The renormalization group methods of Shamir \cite{SHAMIR} seem to me the best way to
make progress on this issue.

Assumption 2) is used in \secref{details} to move from degenerate
to nondegenerate masses in  the four-flavor case.
The most important ``obstruction'' here would seem to be the possible existence an essential
singularity in $V[s]$ at $s=0$; I speculate below on how this possibility might be eliminated.
Note that the existence of such singularity 
immediately would imply that \schpt\ is incorrect. A second way the assumption could
be violated would be the presence of a phase boundary at a finite distance from $s=0$.  This would
imply the existence of a region of mass differences in which \schpt\ is valid, and another
region of larger mass differences in which \schpt\ is invalid.  Generically,  I would expect
an abrupt change like this to cause significant effects that would likely have been noticed in
simulations if they occurred within the parameter ranges studied. 
Both types of potential analyticity violations certainly merit further investigation, however.

Assumption 3) allows me in \secref{extension} to extend the result in the four-flavor case 
to the more interesting cases with fewer than four light flavors.
It should be possible to test this assumption
numerically by simulating a four-flavor theory in
appropriate mass range  and seeing if it is
describable at low energy by the proper chiral theory with a decoupled
charm quark, $(3,4,\fourth)_\chi$.
Such tests are under consideration by the MILC Collaboration and may be
performed in the near future.  The main uncertainty is the precision at
which these tests could be made, which would strongly influence how
convincing they would be.

\bigskip
\bigskip
To investigate a possible essential singularity,
I restrict myself to diagonal  sources, constant in space-time. In other
words, I consider a function $V$ of the four mass differences 
from the degenerate mass $\bar m$.  To correspond with the previous notation, I
write $V=V(\hat s)$, with $\hat s^{ij} \equiv \delta^{ij} \epsilon_j$ and $\epsilon_j = m_j-\bar m$.
Considering $\hat s$ to be complex, the arguments in \secref{expansion} still go through
formally, although one may want to put the system in finite volume
to avoid any dangers from $\int d^4x$ with constant sources.  We then have a complex function
$V(\hat s)$, all 
of whose complex derivatives vanish at $\hat s=0$.  
This would forbid essential
singularities in $V(\hat s)$, which do not have well-defined complex derivatives. 

What would be needed to make such an argument reasonably rigorous?  
On the \schpt\ side, we are defining $(4,4,\fourth)_\chi$
in (chiral) perturbation theory, so I do not expect problems at any finite
order in adding
small, complex $\epsilon_j$ to the masses in Euclidean space.
On the other hand, we do not know what
$(4,``1")_\chi$ is {\it a priori}\/, so we would need to add $\epsilon_j$ to the masses
in the QCD-level theory $(4,``1")_{LQCD}$.  
The main problem there
seems to be that a complex $\hat s$ makes the determinant complex. The issue
of how one chooses the phase of the fourth root thus becomes relevant, as it
is for the case of nonzero chemical potential \cite{CHEMICAL-POTENTIAL}. Unlike
the chemical potential case, however, the imaginary
part of $\epsilon_j$ adds a constant amount to all eigenvalues of flavor $j$.
Furthermore,  $\epsilon_j$   may be taken very small, \ie much less 
than both $\bar m$ and $\Lambda_{QCD}$.
I am hopeful therefore that any phase ambiguities can be shown to be manageable,
but that remains to be seen.  
A further difficulty could come in trying to  ``match'' $(4,``1")_{LQCD}$ onto 
 $(4,4,\fourth)_\chi$ in order to turn statements about smoothness of each theory
separately into statements about $V[s]$.

I conclude with a two additional comments:
\begin{itemize}
\item{} Because the ``wrong'' mesons, which may be lighter than the physical
states, contribute to correlation functions at nonzero lattice spacings in \schpt,
the infinite-distance limit of some quantities may not commute with the 
continuum limit.  This order of limits issue is very similar to
that concerning the chiral and continuum limits, described in Ref.~\cite{LIMITS}.
It is not a practical problem, since of course only finite distances are relevant
to simulations, and the extrapolation can be taken in the proper order with the aid of
\schpt. 

\item{} 
There is nothing in my argument that $(4,4,\fourth)_\chi$ is the
correct chiral theory for four flavors of fourth-rooted staggered quarks
that is really dependent on the fact that we are taking the {\it fourth}\/ root.
The same arguments would also imply that $(3,4,\third)_\chi$ is
the chiral theory for three flavors of staggered quarks for which the
{\it third}\/ root is taken!  The decoupling arguments in
\secref{extension} (which presumably still apply)  would say further
that  $(n_F,4,\third)_\chi$ gives the chiral theory for $n_F<3$ flavors
of third-rooted staggered fields.   There is no contradiction here:
$(n_F,4,\third)_\chi$ describes a sick theory,
{\it even in the continuum limit}\/, except for the uninteresting
case of 3 degenerate flavors.  Since a staggered
field always has four tastes, only a fourth (or square) root
the root can describe an integer number of flavors (and therefore a local
action) in the continuum limit.
The $n_F=1$ example from \secref{one-flavor} provides a simple illustration:
the contributions from light pions in the second and third lines of \eq{Gq}
vanish in the continuum if and only if $n_R=R=1/4$.  (I ignore the trivial case $R=0$, as well as
$n_R=R=-1/4$, which violates the spin-statistics theorem.)
Even $n_R=R=1/2$ leaves some light-pion
contributions, as it should since that is really a two-flavor case. 

\end{itemize}

\section*{ACKNOWLEDGMENTS}
I am very grateful to Maarten Golterman and Yigal Shamir for extensive discussions,
invaluable suggestions, and careful readings of the manuscript,
as well as for a question by Maarten on whether the rooted staggered case should be considered
a mixed theory, which started me thinking along the present lines.
I also thank Jon Bailey, 
Carleton DeTar, Urs Heller, Andreas Kronfeld, as well as other MILC and
Fermilab colleagues, for helpful comments and questions. Finally, I am indebted
to Carl Bender for discussions on analyticity and high orders of perturbation theory.
This work is
partially supported by the US Department of Energy under grant
DE-FG02-91ER40628.

\vfill\eject

\end{document}